\documentclass[journal]{IEEEtran}
\ifCLASSINFOpdf
  \usepackage[pdftex]{graphicx}
\else
  \usepackage[dvips]{graphicx}
\fi
\usepackage{amsmath}
\usepackage{caption}
\usepackage{amssymb}
\usepackage{bm}
\usepackage{upgreek}
\usepackage{tabularx}
\usepackage{float}
\usepackage{xcolor}
\usepackage[normalem]{ulem}
\usepackage{enumitem}

\usepackage{todonotes}

\usepackage{booktabs} % used for creating tables
\usepackage{algorithm} % used for describing algorithms
\usepackage{algorithmic}
\usepackage{multirow}
\usepackage{svg}
\usepackage[switch, pagewise]{lineno}
\usepackage{soul}
\usepackage{epstopdf}
\usepackage{physics}
\AppendGraphicsExtensions{.tif}
\usepackage{placeins}
\usepackage{graphicx} % resize table

\begin{document}
\title{HyperLISTA-ABT: An Ultra-light Unfolded Network for Accurate Multi-component Differential Tomographic SAR Inversion}

\author{Kun~Qian,
        Yuanyuan~Wang,~\IEEEmembership{Member,~IEEE;}
        Peter~Jung,~\IEEEmembership{Member,~IEEE;}
        Yilei~Shi, ~\IEEEmembership{Member,~IEEE;}
        and~Xiao~Xiang~Zhu,~\IEEEmembership{Fellow,~IEEE}% <-this % stops a space
\thanks{This work is supported by the European Research Council with the grant agreement No. [ERC-2016-StG-714087], Acronym: So2Sat, by the Helmholtz Association through the Framework of the Helmholtz Excellent Professorship ``Data Science in Earth Observation - Big Data Fusion for Urban Research''(grant number: W2-W3-100), and by the German Federal Ministry of Education and Research in the framework of the international future AI lab "AI4EO -- Artificial Intelligence for Earth Observation: Reasoning, Uncertainties, Ethics and Beyond" (grant number: 01DD20001).}
\thanks{\emph{Corresponding author: Xiao Xiang Zhu.}}
\thanks{K. Qian, Y. Wang, P. Jung and X. X. Zhu are with the Chair Data Science in Earth Observation, Technical University of Munich, Munich, Germany. (e-mails: \{kun9361.qian,peter.jung,y.wang,xiaoxiang.zhu\}@tum.de).}% <-this % stops a space
%\thanks{Y. Wang and X. Zhu are also with the Department of EO Data Science, Remote Sensing Technology Institute, German Aerospace Center, Oberpfaffenhofen, Germany.}% <-this % stops a space
\thanks{Y. Shi is with the Chair of Remote Sensing Technology, Technical University of Munich, Munich, Germany. (e-mail: yilei.shi@tum.de)}% <-this % stops a 
}


\markboth{Journal of \LaTeX\ Class Files,~Vol.~14, No.~8, August~2015}%
{Shell \MakeLowercase{\textit{et al.}}: Bare Demo of IEEEtran.cls for IEEE Journals}



\maketitle

\begin{abstract}
Deep neural networks based on unrolled iterative algorithms have achieved remarkable success in sparse reconstruction applications, such as synthetic aperture radar (SAR) tomographic inversion (TomoSAR). However, the currently available deep learning-based TomoSAR algorithms are limited to three-dimensional (3D) reconstruction. The extension of deep learning-based algorithms to four-dimensional (4D) imaging, i.e., differential TomoSAR (D-TomoSAR) applications, is impeded mainly due to the high-dimensional weight matrices required by the network designed for D-TomoSAR inversion, which typically contain millions of freely trainable parameters. Learning such huge number of weights requires an enormous number of training samples, resulting in a large memory burden and excessive time consumption. To tackle this issue, we propose an efficient and accurate algorithm called HyperLISTA-ABT. The weights in HyperLISTA-ABT are determined in an analytical way according to a minimum coherence criterion, trimming the model down to an ultra-light one with only three hyperparameters. Additionally, HyperLISTA-ABT improves the global thresholding by utilizing an adaptive blockwise thresholding scheme, which applies block-coordinate techniques and conducts thresholding in local blocks, so that weak expressions and local features can be retained in the shrinkage step layer by layer. Simulations were performed and demonstrated the effectiveness of our approach, showing that HyperLISTA-ABT achieves superior computational efficiency and with no significant performance degradation compared to state-of-the-art methods. Real data experiments showed that a high-quality 4D point cloud could be reconstructed over a large area by the proposed HyperLISTA-ABT with affordable computational resources and in a fast time.

\end{abstract}

% \begin{abstract}

% \end{abstract}


% \fi
\IEEEpeerreviewmaketitle

\section{Introduction}
Synthetic aperture radar tomography (TomoSAR) has attracted significant interest due to its capability in 3-D reconstruction, particularly for urban areas \cite{Zhu2010Very} \cite{tomosar_urban1} \cite{tomosar_urban2} \cite{tomosar_urban3}. Compressive sensing \cite{CS1} \cite{CS2} (CS)-based algorithms are usually preferred for solving TomoSAR inversion \cite{Zhu2010Tomographic} \cite{Shi2018Fast} \cite{zhu_joint}. However, the heavy computational cost of CS-based methods makes them less applicable for large-scale processing. Among the different methods aiming to tackle this issue, deep neural networks have been employed in speeding up TomoSAR inversion. In the work presented in \cite{dl_tomosar1}, TomoSAR inversion was approached as a classification problem, and a conventional convolutional neural network (CNN) was employed to solve the problem. However, this approach was limited to the detection of single scatterers, and it did not fully address the challenges of TomoSAR inversion for complex scenes with multiple scatterers and variations in the elevation direction. More recently, thanks to an emerging deep learning technique called deep unfolding \cite{deep_unfolding}, the authors proposed $\boldsymbol{\gamma}$-Net in \cite{gamma-net_QK} for improving the unrolled iterative shrinkage thresholding algorithm (ISTA)-network. It was shown that $\boldsymbol{\gamma}$-Net could succeed in accelerating the processing speed by 2-3 order of magnitude while maintaining a comparable super-resolution power and location accuracy compared to second-order CS-solvers. In addition, a gated recurrent structure, dubbed as complex-valued sparse minimal gated units (CV-SMGUs), was proposed in \cite{SMGU_QK} that incorporates historical information into the dynamics of network, thus preserving the full information. As discussed in \cite{SMGU_QK}, CV-SMGUs could outperform $\boldsymbol{\gamma}$-Net by a fair margin. 

However, to the best of our knowledge, deep learning-based TomoSAR algorithms are to date still confined to 3-D reconstruction cases. Considering that spaceborne datasets are usually acquired in the repeat-pass mode at different time stamps, often over several years, it is necessary to additionally account for a potential deformation of objects in the estimation, such as seasonal motion caused by thermal dilation or linear motions like subsidence. The 4-D imaging technique taking into account additional deformation parameters is known as differential TomoSAR (D-TomoSAR) \cite{Zhu2010Very} \cite{four-D} \cite{D-TomoSAR1} \cite{zhu_4D_tomosar}.

The limitation of deep learning-based algorithms in solving D-TomoSAR inversion is mainly attributed to the high-dimensional weight matrices to be learned in the network. For modern deep learning-based algorithms, like $\boldsymbol{\gamma}$-Net and CV-SMGUs, the size of the weight matrices is usually related to the discretization level. In D-TomoSAR cases, especially when multi-component motion terms are considered, weight matrices can easily contain over one million free trainable parameters. As a consequence, it would be extremely computationally inefficient to learn those weights without mentioning the enormous number of training samples required. A detailed analysis of this issue can be found in Section III.A of the present paper.

% To cope with the cumbersome learning of huge weights when deep learning-based algorithms are adopted, an analytic weight determination method was firstly proposed in \cite{liu2018alista} and further developed as HyperLISTA \cite{HyperLISTA}. The analytic weight determination provides us a novel insight in applying deep learning-based algorithms in D-TomoSAR inversion that the weights can be optimized in a data-free approach, thus overcoming the issued limitation. 
To tackle the computational challenges posed by learning huge weights, a pioneering solution was introduced in the seminal work by Liu et al. \cite{liu2018alista}. In their research, they proposed an analytical weight determination method. This method was further refined by them and extended into HyperLISTA \cite{HyperLISTA}. By employing analytical weight determination, a novel perspective emerges for leveraging deep learning-based algorithms in D-TomoSAR inversion. Specifically, the optimization of weights using a data-free approach becomes possible, circumventing the need for an extensive number of training samples. This innovative approach effectively overcomes the limitations that were previously encountered in deep learning-based D-TomoSAR inversion.

However, it is important to note that, similar to LISTA, HyperLISTA employs a global thresholding scheme where a unified threshold is used to prune all entries. The choice of an appropriate threshold is thus crucial. A high threshold value may result in the loss of significant information \cite{SMGU_QK}, such as local features generated by echoed signals from dark scatterers. On the other hand, a low threshold value can delay convergence and yield a solution that lacks sufficient sparsity.

\subsection{Contribution of the present study to the field}
%  and benefit from the efficiency provided by deep learning, we proposed an ultra-light model, named HyperLISTA-ABT, by improving HyperLISTA \cite{HyperLISTA} with an adaptive thresholding scheme. In HyperLISTA-ABT, the weights can be determined analytically using a minimum coherence criteria that a system matrix with low mutual coherence implies a recovery of high probability. 

% However, the same as LISTA, HyperLISTA employs a global thresholding scheme, in which a unified threshold is adopted to prune all entries. It is usually critical to choose a proper threshold. A threshold of large value may cause loss of significant information \cite{SMGU_QK}, e.g. some local features caused by echoed signal from dark scatterers may get discarded. A threshold of small value delays the convergence and makes the solution not sparse enough.
To overcome the aforementioned issue in to D-TomoSAR inversion, we proposed an ultra-light model, named HyperLISTA-ABT, that improves HyperLISTA \cite{HyperLISTA} through incorporating an adaptive blockwise thresholding (ABT) scheme. Same as HyperLISTA, the proposed HyperLISTA-ABT can be viewed as an unrolled ISTA network, whereas the weight matrices therein can be determined with analytical optimization according to the minimum coherence criterion. A system matrix with low mutual coherence implies a recovery of high probability, which is the fundamental concept of compressive sensing. The adaptive blockwise thresholding scheme in HyperLISTA-ABT enables updating the block coordinates and conducting a shrinkage in local regions. Moreover, the blocksize is adjusted layer by layer for a better fine-focusing ability. The main contribution of this paper is listed as follows:
\begin{enumerate}
    \item We propose the efficient and accurate algorithm HyperLISTA-ABT and, to the best of our knowledge, are the first to apply deep neural networks to solve D-TomoSAR and multi-component D-TomoSAR inversion.
    \item We apply a block-coordinate technique and propose an adaptive blockwise thresholding scheme to replace global thresholding in most shrinkage thresholding methods. Therefore, the local features from a weakly echoed signal can be possibly retained.
    \item We carry out a systematic performance evaluation using both simulated and real data. The results demonstrate that the proposed HyperLISTA-ABT provides competitive estimation accuracy and superior computational efficiency. Large-scale D-TomoSAR processing was conducted, demonstrated by a 4-D point cloud reconstruction over Las Vegas.
\end{enumerate}

\section{Background}
\subsection{High-dimensional SAR imaging model for D-TomoSAR}
\begin{figure}[h]
    \centering
    \includegraphics[width=0.49\textwidth]{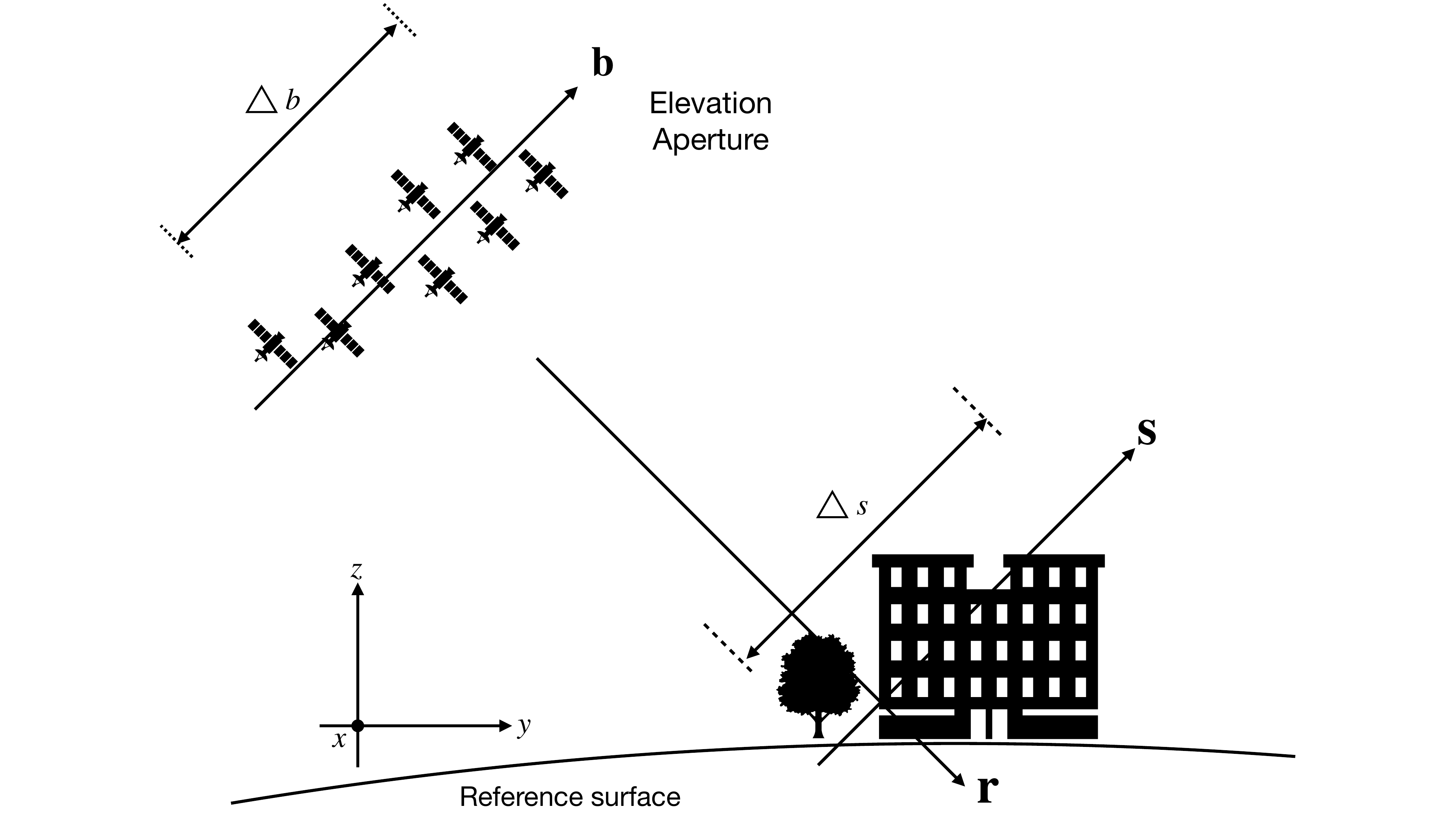}
    \caption{SAR imaging geometry at a fixed azimuth position. The elevation synthetic aperture is built up by acquisition from slightly different incidence angles. Flight direction is orthogonal into the plane.}
    \label{fig:tomosar}
\end{figure}
D-TomoSAR, employs multi-baseline and multi-temporal SAR acquisitions to estimate scatterering profiles. Based on the estimated scatterering profiles, we can reconstruct the 3-D distribution of scatterers along the elevation direction and the motion history assigned to each elevation position \cite{Zhu2010Very}\cite{D-TomoSAR1}\cite{D-TomoSAR2}. The following describes the D-TomoSAR imaging model:
\begin{equation}
    g_n=\int_{\Delta s} \boldsymbol{\gamma}(s) \exp \left(-j 2 \pi\left(\xi_n s+2 d\left(s, t_n\right) / \lambda\right)\right) \mathrm{d} s.
    \label{eq:sin_motion}
\end{equation}
where $g_n$ is the complex-valued SAR acquisition at a certain azimuth-range pixel at time $t_n$ ($n=1,2,\cdots,N$); $\boldsymbol{\gamma}(s)$ denotes the scatterering profile along the elevation direction with an extent of $\triangle s$; $\xi = 2b_n / \lambda r$ is the elevation frequency proportional to the respective aperture position $b_n$; $d(s,t_n)$ depicts the line-of-sight (LOS) motion, which is a function of elevation and time. The LOS motion relative to the master acquisition can be modeled with a linear combination of M base functions $\tau_M(t_n)$
\begin{equation}
    d\left(s, t_n\right)=\sum_{m=1}^M p_m(s) \tau_m\left(t_n\right)
\end{equation}
where $p_m(s)$ is the corresponding motion coefficient to be estimated. The choice of the base functions $\tau_m(t_n)$ depends on the underlying physical motion processes. Great details about how to choose proper base functions can be found in \cite{zhu_4D_tomosar}. Taking multi-component motion into consideration, we generalize Eq. (\ref{eq:sin_motion}) as:
\begin{align}
 g_n &=\int \ldots \iint \gamma(s) \delta\left(p_1-p_1(s), \ldots, p_M-p_M(s)\right) \\ \nonumber
&  \exp \left(j 2 \pi\left(\xi_n s+\eta_{1, n} p_1+\ldots+\eta_{M, n} p_M\right)\right) d s d p_1 \ldots d p_M
\label{eq:mul_motion}
\end{align}
The inversion of the system model with multi-component motion retrieves the elevation information as well as the the motion history assigned to each elevation position, even if multiple scatterers are overlaid inside an resolution unit. Therefore, we can acquire a high-dimensional map of the scatterers. In the presence of noise $\boldsymbol{\varepsilon}$, the discrete high-dimensional D-TomoSAR model can be expressed as:
\begin{equation}
    \mathbf{g} = \mathbf{R} \boldsymbol{\gamma} + \boldsymbol{\varepsilon}
    \label{eq:tomosar}
\end{equation}
where $\mathbf{g} \in \mathbb{C}^{N \times 1}$ is the complex-valued SAR measurement vector and $\mathbf{R} \in \mathbb{C}^{N \times L}$ is the irregular sampled Fourier transformation steering matrix, where $N$ is the number of SAR acquisitions and $L$ is the amount of the discretization in the signal to be reconstructed.

As investigated in \cite{Zhu2010Tomographic}, usually only a few (less than 4) scatterers are overlaid inside an individual pixel in urban areas, such that $\boldsymbol{\gamma}$ is sufficiently sparse so that retrieving $\boldsymbol{\gamma}$ can be formulated as a sparse reconstruction problem. Accordingly, solving $\boldsymbol{\gamma}$ in the presence of noise can be formulated as a basis pursuit denoising (BPDN) optimization problem, which can be expressed as follows:
\begin{equation} \label{gamma estimate}
    \hat{\boldsymbol{\gamma}}=\arg \min _{\boldsymbol{\gamma}}\left\{\|\mathbf{g}-\mathbf{R} \boldsymbol{\gamma}\|_{2}^{2}+\lambda \|\boldsymbol{\gamma}\|_{1}\right\},
\end{equation}
where $\lambda$ is the regularization parameter controlling the data-fit terms and the signal sparsity. Great details about how to choose a proper value of $\lambda$ according to the noise level can be found in \cite{BPDN}.

\subsection{Review of the deep learning-based TomoSAR algorithms and their limitation in solving D-TomoSAR inversion}
\begin{figure}[H]
    \centering
    \includegraphics[width=0.98\linewidth]{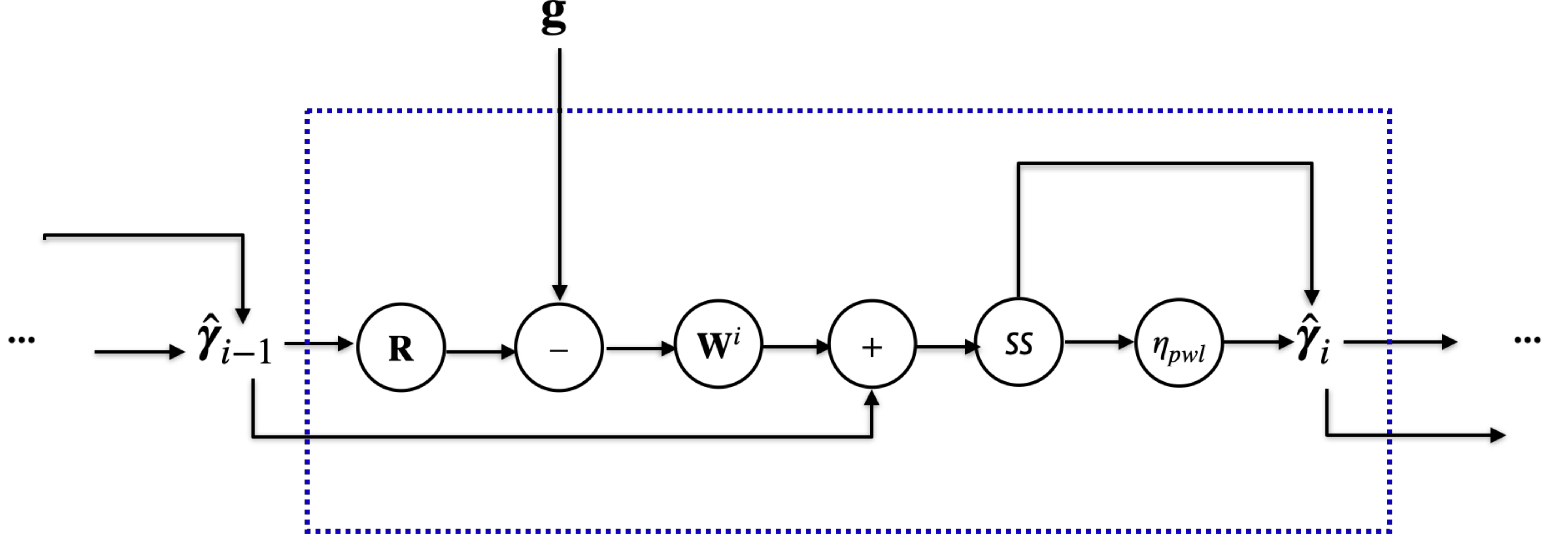}
    \caption{Illustration of an intermediate layer in $\boldsymbol{\gamma}$-Net.}
    \label{fig:layer_gamma_net}
\end{figure}
In \cite{gamma-net_QK} \cite{SMGU_QK}, the respective authors proposed two advanced deep learning-based algorithms by improving unrolled neural networks. Experimental results on both laboratory and real data demonstrated their strong super-resolution power and high location accuracy. However, their application to date is still limited to 3-D reconstruction. Taking $\boldsymbol{\gamma}$-Net as an example, we will explain the difficulty of applying deep learning-based algorithms for solving D-TomoSAR inversion. To start with, we briefly go through the basics of the $\boldsymbol{\gamma}$-Net architecture. Fig. \ref{fig:layer_gamma_net} illustrates us the structure of an intermediate layer of $\boldsymbol{\gamma}$-Net, which can be formally defined as follows:
\begin{equation}
     \hat{\boldsymbol{\gamma}}_k = {\eta}_{\theta_k}^{\rho^k} \{\hat{\boldsymbol{\gamma}}_{k-1} + \mathbf{W}^{H}_{k} (\mathbf{g} - \mathbf{R} \hat{\boldsymbol{\gamma}}_{k-1}), \boldsymbol{\theta_i}\}
     \label{eq:lista_cpss}
\end{equation}
More details about $\boldsymbol{\gamma}$-Net can be found in \cite{gamma-net_QK}.
% \textbf{\textit{SS}} indicates a special thresholding scheme called support selection and it selects $\rho^i$ percentage of entries with the largest magnitude. Then the selected part will be directly fed to the next layer, bypassing the current shrinkage step. More details about $\boldsymbol{\gamma}$-Net formulation can be found in \cite{gamma-net_QK}.

As we can see, in each $\boldsymbol{\gamma}$-Net layer, a weight matrix $\mathbf{W}_{k}$ of the size $N \times L$ needs to be learned. For 3-D reconstruction cases, the value of $L$ is only determined by the grids number after the discretization of the elevation extent, thus it is typically in the range of hundreds for spaceborne sensors and $N \times L$ will be in thousands then. However, this number increases exponentially in D-TomoSAR cases when multi-component motion terms, usually linear and periodic motions, are taken into consideration. The training of the network then becomes conversely a challenge due to the tremendous amount of free trainable parameters. For instance, when we consider two motion terms, i.e., linear and seasonal motion, the value of $L$ will be determined by the product of the discretized grid numbers along each direction $L = L_s \times L_v \times L_a$, where $L_s$, $L_v$, and $L_a$ indicate the discretization levels for elevation, linear motion, and seasonal motion, respectively. A very conservative level of discretization in elevation, linear motion, and periodic motion for TerraSAR-X image stacks $L_s$, $L_v$, and $L_a$ would be 200, 50, and 50. When multiplied, the value of $L$ will then be 0.5 million, meaning that, there will be millions of parameters to be learned in each weight matrix. Such large weight matrices result in two unavoidable downsides. First, the model tends to converge at the ground truth instead of the LASSO minimizer, because the update direction $\mathbf{W}^{H}(\mathbf{g-R\gamma})$ does not align with the gradient of the $l_2$ term in the LASSO objective $\mathbf{R}^{H}(\mathbf{g-R\gamma})$. Therefore, we always need to train the model in a supervised way. Consequently, a massive number of training samples are required to train the model with huge weight matrices, thus making the training procedure extremely inefficient. Second, the training of the huge model requires a significant amount of GPU memory, which is usually not feasible with consumer-level GPUs.

\section{Methodology}
\subsection{HyperLISTA with adaptive blockwise thresholding (HyperLISTA-ABT)}
To circumvent the tedious and troublesome model training caused by needing to learn huge weight matrices, an analytical weight optimization method, which is based on coherence minimization, was proposed in ALISTA \cite{liu2018alista} to determine the weights in an unrolled neural network designed for sparse recovery, such as LISTA. ALISTA combines the superior empirical performance of fully learned methods and significantly reduces the number of parameters, leaving only thresholds and stepsize parameters to be learned. In addition, an ultra-light model, called HyperLISTA, was proposed in \cite{HyperLISTA}, which further trimmed down the training complexity. In HyperLISTA, weight matrices can be computed in a similar way to \cite{liu2018alista} and the training is reduced to tuning only three hyperparameters from the data. The following shows us the formal update rules of HyperLISTA:
\begin{align}
    \boldsymbol{\gamma}_{k+1}=\eta_{\theta^{k}}^{p^{k}}(\boldsymbol{\gamma}_{k}+ \boldsymbol{W}^H(\boldsymbol{g}-\boldsymbol{R} \boldsymbol{\gamma}_{k}) \\ \nonumber +\beta^{k}(\boldsymbol{\gamma}_{k}-\boldsymbol{\gamma}_{k-1}))
\end{align}
where
\begin{align}
& \theta^{k}=c_1  \left\|\boldsymbol{R}^{+}\left(\boldsymbol{R} \boldsymbol{\gamma}_{k}-\boldsymbol{g}\right)\right\|_1 \\
& \beta^{k}=c_2 \left\|\boldsymbol{\gamma}_{k}\right\|_0 \\
& p^{k}=c_3 \min \left(\log \left(\frac{\left\|\boldsymbol{R}^{+} \boldsymbol{g}\right\|_1}{\left\|\boldsymbol{R}^{+}\left(\boldsymbol{R} \boldsymbol{\gamma}_{k}-\boldsymbol{g}\right)\right\|_1}\right), L\right)
\end{align}
where $c_1, c_2$, and $c_3$ indicate the three hyperparameters to be tuned. It is possible to learn the hyperparameters via backpropagation, albeit this method may be an overkill as it involves passing gradients through deep neural network layers to learn just three parameters. Less computationally expensive methods, such as grid search, could be employed to obtain a set of proper hyperparameters. Despite providing a bit less accurate estimate, the empirical findings in \cite{HyperLISTA} showed that HyperLISTA is robust to perturbations in the values of c1, c2, and c3. In grid search, a coarse grid is first applied to find an interest region, and then this is zoom-in with a fine-grained grid. The hyperparameters are determined by minimizing the normalized mean square error (NMSE) over the simulated ground truth. The NMSE is defined as:
\begin{equation}
    \mathrm{N M S E}=\frac{1}{T} \sum \frac{\|\hat{\boldsymbol{\gamma}}-\boldsymbol{\gamma}\|_{2}^{2}}{\|\boldsymbol{\gamma}\|_{2}^{2}}
\end{equation}
where T denotes the number of samples, and $\eta_{\theta^{(k)}}^{p^{(k)}}$ is the soft-thresholding function combined with the support selection scheme,
\begin{equation}
    {\eta}_{\theta_{(k)}}^{\rho^{(k)}}(\boldsymbol{\gamma}^{(k)})=\left \{
    \begin{array}{lr}
         \boldsymbol{\gamma}^{(k)}  & i \in \mathcal{S}^{\rho^{(k)}}(\boldsymbol{\gamma}) \\
         \eta_{st}(\boldsymbol{\gamma}^{(k)}, \theta_{(k)})  & i \notin \mathcal{S}^{\rho^{(k)}}(\boldsymbol{\gamma})
    \end{array}
    \right. .
    \label{eq:ss}
\end{equation}
$\mathcal{S}^{\rho^{(k)}}(\boldsymbol{\gamma})$ contains the entries with the $\rho^{(k)}$ largest magnitudes. $\mathbf{W}$ denotes the optimized weight matrix determined with the minimum coherence criterion, which is defined as follows:
\begin{align}
\nonumber
    \hat{\mathbf{W}} &= \arg \min_{\mathbf{W}} \mu(\mathbf{W}, \mathbf{R}) \\ \nonumber 
    &= \arg \min_{\mathbf{W}} \inf _{\mathbf{W} \in \mathbb{C}_{N \times L}}  \max _{i \neq j} \mathbf{W}_{:, i}^T \mathbf{R}_{:, j} \\
    &\text { s.t. } \forall i \in\{1, \ldots, L\}: \mathbf{W}_{:, i}^T \mathbf{R}_{:, i}=1
    \label{eq:min_coh}
\end{align}
Rigorous proof of the convergence and recovery upper and lower bound of HyperLISTA can be found in \cite{HyperLISTA}. An efficient numerical algorithm to calculate the optimized weights is discussed in the Appendix.

Inspired by the outstanding efficiency and performance demonstrated in \cite{HyperLISTA}, we consider HyperLISTA should have great potential in our high-dimensional D-TomoSAR inversion. However, through experiments, we discovered a drawback of HyperLISTA when applied to TomoSAR. Similar to most thresholding algorithms, HyperLISTA suffers from an inherent limitation caused by the global thresholding scheme. Precisely, in the signal projection process for identifying the presence of a dictionary atom within the signal, the selection of an appropriate threshold is of utmost importance. The threshold should be chosen carefully to account for both strong and weak spikes in the reflectivity profile. By selecting a well-suited threshold, the signal projection can distinguish between significant spikes and noise, enabling an accurate identification of dictionary atoms within the signal. However, when utilizing HyperLISTA and other methods that employ global thresholding, the task of selecting an optimal threshold becomes exceedingly challenging. The use of a global threshold implies that the same threshold value is applied uniformly across all entries in the signal. This approach may lead to suboptimal results, as a threshold that effectively captures strong spikes might inadvertently suppress weaker but still meaningful spikes in the reflectivity profile. Consequently, we usually need to choose a relatively small $c_1$ to have a small threshold so that we can maintain some small spikes caused by reflection from dark scatterers. Otherwise, the information of dark scatterers would be discarded in the thresholding step layer by layer. However, the use of a small threshold brings about two main problems. First, the convergence would be considerably slow. Second, small thresholds yield solutions that are not sparse enough. 

To cope with the aforementioned issue and better leverage the power of HyperLISTA in our application, we propose HyperLISTA-ABT, which is an improvement of the original HyperLISTA by incorporation of an adaptive blockwise thresholding (ABT) scheme that explores a local thresholding strategy. The advantages of HyperLISTA-ABT is three-fold. First, it conducts the thresholding in each local block, thus allowing for a more refined thresholding process and possibly retaining weak expressions of reflections from dark scatterers. Then, it becomes possible to better capture the diverse range of spike magnitudes encountered in the signal, enhancing the accuracy and reliability of the reflectivity profile characterization. Second, HyperLISA-ABT has been shown to be more efficient since it updates only one block of variables at each time instead of updating all the variables together. Therefore, HyperLISTA-ABT has been found to be more appropriate for our large-scale and high-dimensional application. Last but not the least, HyperLISTA-ABT reduces the blocksize layerwise and contributes to a better fine-focusing ability. 

According to \cite{block_1} \cite{block_2}, the update rules of HyperLISTA-ABT after applying block coordinate techniques can be written as:
\begin{align}
    \boldsymbol{\gamma}^{k+1}_{i_p}=\eta_{\theta^{k}_{i_p}}(\boldsymbol{\gamma}^{k}_{i_p}+ \boldsymbol{W}_{i_p}^T(\boldsymbol{y}-\boldsymbol{R}_{i_p} \boldsymbol{\gamma}^{k}_{i_p}) \\ \nonumber +\beta^{k}_{i_p}(\boldsymbol{\gamma}^{k}_{i_p}-\boldsymbol{\gamma}^{k-1}_{i_p}))
\end{align}
where $i_p$ is the index of the updated block. To clarify, in HyperLISTA-ABT, we remove the support selection scheme and just use the conventional soft-thresholding function. The threshold $\theta^{(k)}$ and the factor $\beta^{(k)}$ are determined for each block as well:
\begin{align}
& \theta^{k}_{i_p}=c_1  \left\|\boldsymbol{R}_{i_p}^{+}\left(\boldsymbol{R}_{i_p} \boldsymbol{\gamma}^{k}_{i_p}-\boldsymbol{g}\right)\right\|_1 \\
& \beta^{k}_{i_p}=c_2 \left\|\boldsymbol{\gamma}^{k}_{i_p}\right\|_0
\end{align}
where $c_1 >0$, $c_2 >0$, and $c_3 \in (0,1)$ are the three hyperparameters. Notably, $c_3$ is a latent hyperparameter and plays a crucial role in controlling the blocksize despite it not explicitly appearing in the formula. In our application, we usually initialize the blocksize according to the grid number within half of the Rayleigh resolution. The block is chosen with a random variants scheme where $i_p$ follows the probability distribution given by:
\begin{equation}
    P_{i_p}=\frac{L_{i_p}}{\sum_{j=1}^J L_{i_p}}, \quad i_p=1, \ldots, J
\end{equation}
where $J$ is the number of blocks and $L_{i_p} = ||R_{i_p}^T R_{i_p}||$. All the hyperparameters $c_1, c_2$, and $c_3$ can be selected using the same grid search method as in HyperLISTA. 

With the blockwise threshlding scheme, local features and weak expressions can be possibly retained. This is due to the fact that many elements of the entries are not strictly driven to zero but to some extremely small value, thus making the output not strictly sparse. Therefore, a post-processing is usually required to clean the output and make it sparse. The framework of the proposed HyperLISTA-ABT is summarized in the following table.
\begin{algorithm}[h]
\caption{Summary of the proposed algorithm}
\label{table:algorithm}
\begin{algorithmic}
    \STATE \textbf{Generate steering matrix R for given baselines} \\
         \STATE \qquad  \textbf{Analytic} weight optimization $\mathbf{W}$ according to Eq. (\ref{eq:min_coh}) \\
    \STATE \textbf{Tuning of hyperparameters}
         \STATE \qquad  \textbf{Simulate} ground truth of reflectivity profile $\boldsymbol{\gamma}$ \cite{gamma-net_QK} \\
         \STATE \qquad \textbf{Simulate} noise-free SAR acquisitions $\mathbf{g = R} \boldsymbol{\gamma}$\\
         \STATE \qquad \textbf{Grid search} to determine the hyperparameters by \\
         \qquad minimizing NMSE over simulated data \\
     \STATE \textbf{Inference} \\
     \STATE \qquad \textbf{Init:} $ \boldsymbol{\gamma}^{(0)} = \mathbf{R}^H \mathbf{g}$ and blocksize $B_1$ \\
     \STATE \qquad \textbf{for} $k = 1,2, \cdots, K$ \textbf{do} \\
     \STATE \qquad \qquad \textbf{Determine} the number of blocks $J_k$ \\
                \qquad \qquad based on the blocksize $B_k$
     \STATE \qquad \qquad \textbf{for} $i_p = 1,2, \cdots, J_k$ \textbf{do} \\
            \begin{align} \nonumber
 \qquad \qquad  \boldsymbol{\gamma}^{(k+1)}_{i_p}&=\eta_{\theta^{(k)}_{i_p}}(\boldsymbol{\gamma}^{(k)}_{i_p}+ \boldsymbol{W}_{i_p}^T(\boldsymbol{y}-\boldsymbol{R}_{i_p} \boldsymbol{\gamma}^{(k)}_{i_p}) \\ \nonumber & +\beta^{(k)}_{i_p}(\boldsymbol{\gamma}^{(k)}_{i_p}-\boldsymbol{\gamma}^{(k-1)}_{i_p})) \\ \nonumber
  \theta^{(k)}_{i_p}&=c_1  \left\|\boldsymbol{R}_{i_p}^{+}\left(\boldsymbol{R}_{i_p} \boldsymbol{\gamma}^{(k)}_{i_p}-\boldsymbol{g}\right)\right\|_1 \\ \nonumber
 \beta^{(k)}_{i_p}&=c_2 \left\|\boldsymbol{\gamma}^{(k)}_{i_p}\right\|_0
 \end{align}
    \STATE \qquad \qquad \textbf{end for}
     \STATE \qquad \qquad \textbf{Update} blocksize with $B_{k+1} = c_3 \cdot B_k$ \\
     \STATE \qquad \textbf{end for}
     \STATE \textbf{Output clean-up}
     \STATE \textbf{Model order selection and final estimation}
 \end{algorithmic}
\end{algorithm}

\section{Simulations}
To demonstrate the improvement of the proposed HyperLISTA-ABT to the original HyperLISTA and compare it to the state-of-the-art CS-based and deep learning-based methods, we first conducted experiments based on TomoSAR inversion using simulated data. Since existing deep learning-based algorithms are not feasible to use with D-TomoSAR cases as explained in Section I, we only focused on TomoSAR inversion for 3-D reconstruction in the simulation. 
\subsection{Simulation setup}
In the simulation, we conducted a well-known TomoSAR benchmark test \cite{Zhu2010Very} \cite{Zhu2010Tomographic} using the same simulation settings as used in \cite{gamma-net_QK} \cite{SMGU_QK}. specifically, we simulated an interferometric stack containing 25 baselines that are regularly distributed in the range of -135m to 135m and a two-scatterer mixture in each resolution cell. We used the \textbf{\textit{effective detection rate}}, which is able to simultaneously reflect the super resolution power and elevation estimation accuracy, to fairly evaluate the performance. Detailed definition of the \textbf{\textit{effective detection rate}} can be found in \cite{gamma-net_QK} \cite{SMGU_QK}. 

% For each scatterer overlaid in a resolution cell, the simulation should satisfy the following rules:
% \begin{enumerate}
%     \item The scattering phase $\phi$ follows an uniform distribution in the range of $(-\pi, \pi)$, i.e. $\phi \sim U(-\pi,\pi)$
%     \item The scattering amplitude $A$ follows an uniform distribution in the range of $(1, 4)$, i.e. $A \sim U(1,4)$
%     \item The elevation $s$ follows an uniform distribution in the range of $(-20m, 300m)$, i.e. $s \sim U(-20, 300)$
% \end{enumerate}
% Accordingly, different amplitude ratio, different scattering phase offset as well as different elevation distance between the two scatterers are considered. 

\subsection{Performance improvement compared to the original HyperLISTA}
\begin{figure*}[h!]
    \centering
    \includegraphics[width=0.99\linewidth]{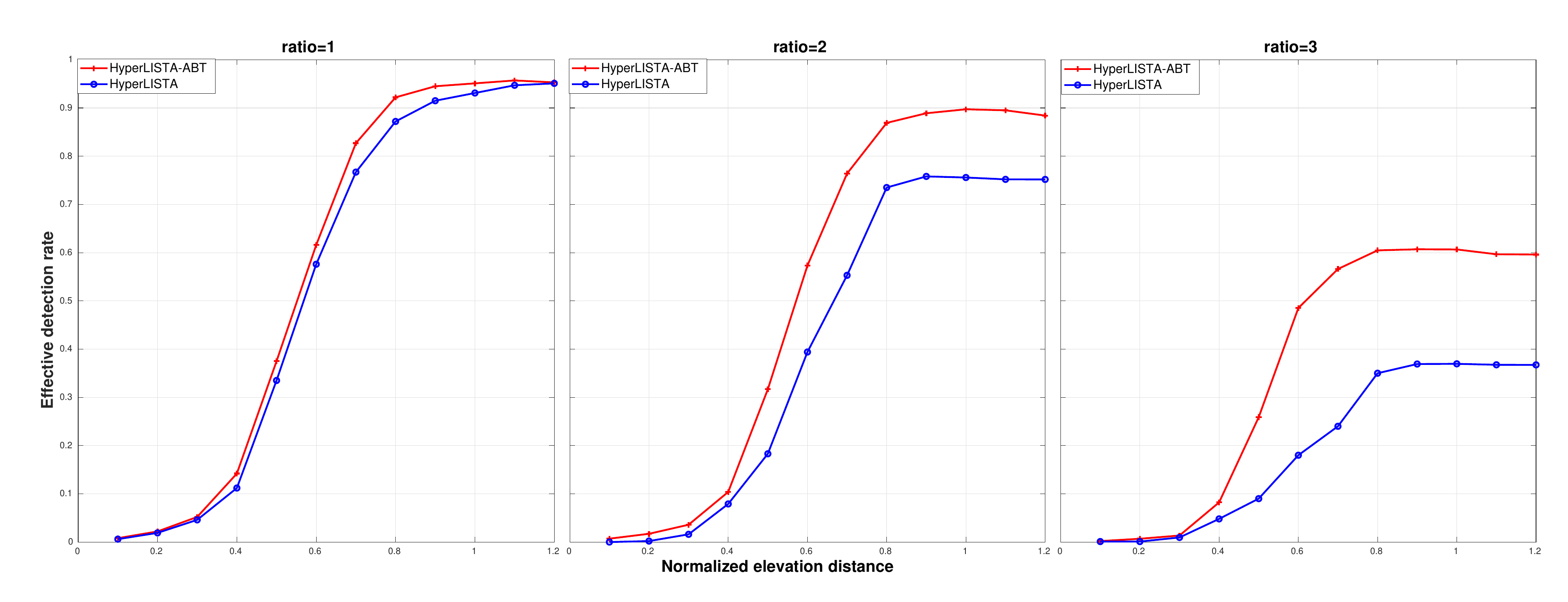}
    \caption{Effective detection rate of HyperLISTA-ABT and the original HyperLISTA with respect to the normalized elevation distance at different amplitude ratios. The overlaid double scatterers were set to have an identical phase and the SNR level was 6dB. HyperLISTA-ABT significantly outperformed HyperLISTA at high amplitude ratios between the scatterers.}
    \label{fig:amp_ratio}
\end{figure*}

The first experiment set out to study the performance improvement of HyperLISTA-ABT compared to the original HyperLISTA. In this experiment, the overlaid double scatterers were simulated to have identical scattering phase varying amplitude ratios. The two algorithms were set to have 15 layers, which is a typical number for a LISTA network and its variants. Fig. \ref{fig:amp_ratio} shows the effective detection rate of HyperLISTA-ABT and the original HyperLISTA as a function of the normalized elevation distance between the simulated double scatterers at 6dB SNR at different amplitude ratios. The results demonstrate that HyperLISTA-ABT achieved a significantly higher effective detection rate than the original HyperLISTA. Both algorithms (in fact, all other methods) experience performance degradation with respect to an increase in amplitude ratio. This is attributed to two main factors. Firstly, dark scatterers experience a large bias in their elevation estimates at high amplitude ratios due to their elevation estimates approaching the more prominent ones. Consequently, many detections of double scatterers will not be recognized as effective due to the large elevation estimate bias. Secondly, the energy of dark scatterers is close to the noise level at high amplitude ratios. This makes it particularly challenging for HyperLISTA, which employs a global thresholding scheme, to detect the local features of dark scatterers. Further elaborating, when high-intensity scatterers are present in the signal, their strong energy can overshadow the low-energy regions where dark scatterers are located. This overshadowing effect can lead to the suppression or even annihilation of the weaker expressions associated with the dark scatterers. Consequently, the presence of these strong intensity scatterers can mask or obscure the signals originating from the dark scatterers, making their detection and characterization challenging. In contrast, HyperLISTA-ABT conducts thresholding in each local block, which can allow retaining local information and, thus, it can detect dark scatterers. This results in a higher effective detection rate at high amplitude ratios.

% As one can see, HyperLISTA-ABT delivers much higher effective detection rate than the original HyperLISTA despite the fact that the performance of the both algorithms degrades with the increase of the amplitude ratio. The performance degradation attributes to two main issues. First, dark scatterers suffer from large bias at high amplitude ratio since their elevation estimates tend to approach the prominent ones. As a consequence, many detection of double scatterers will not be recognized as effective due to large elevation estimate bias. Second, the energy of dark scatterers is close to the noise level at high amplitude ratio. Accordingly, it is difficult for HyperLISTA with the global thresholding sceme to detect the local feature of dark scatterers. In other words, scatterers with prominent intensity can "cast a shadow" over low-energy regions of dark scatterers in the signal and the weak expressions of dark scatterers will be annihilated. In the contrast, HyperLISTA-ABT conducts thresholding in each local block, which is able to possibly retain local information, thus detecting dark scatterers and contributing higher effective detection rate at high amplitude ratio. %

\subsection{Comparison with the state-of-the-art algorithms}
In this section, we compared HyperLISTA-ABT to other state-of-the-art algorithms for further evaluation, which are deep learning-based algorithms $\boldsymbol{\gamma}$-Net \cite{gamma-net_QK} and CV-SMGUs \cite{SMGU_QK}, as well as the traditional CS-based method SL1MMER \cite{Zhu2012Super-Resolution} with second-order optimization. To highlight the super-resolution ability of these methods, we also involved a conventional spectral estimator SVD-Wiener \cite{Zhu2010Very} as a baseline in the comparison.

\begin{figure*}[h!]
	\centering
    \begin{minipage}[t]{0.49\linewidth}
    \includegraphics[width=0.98\textwidth]{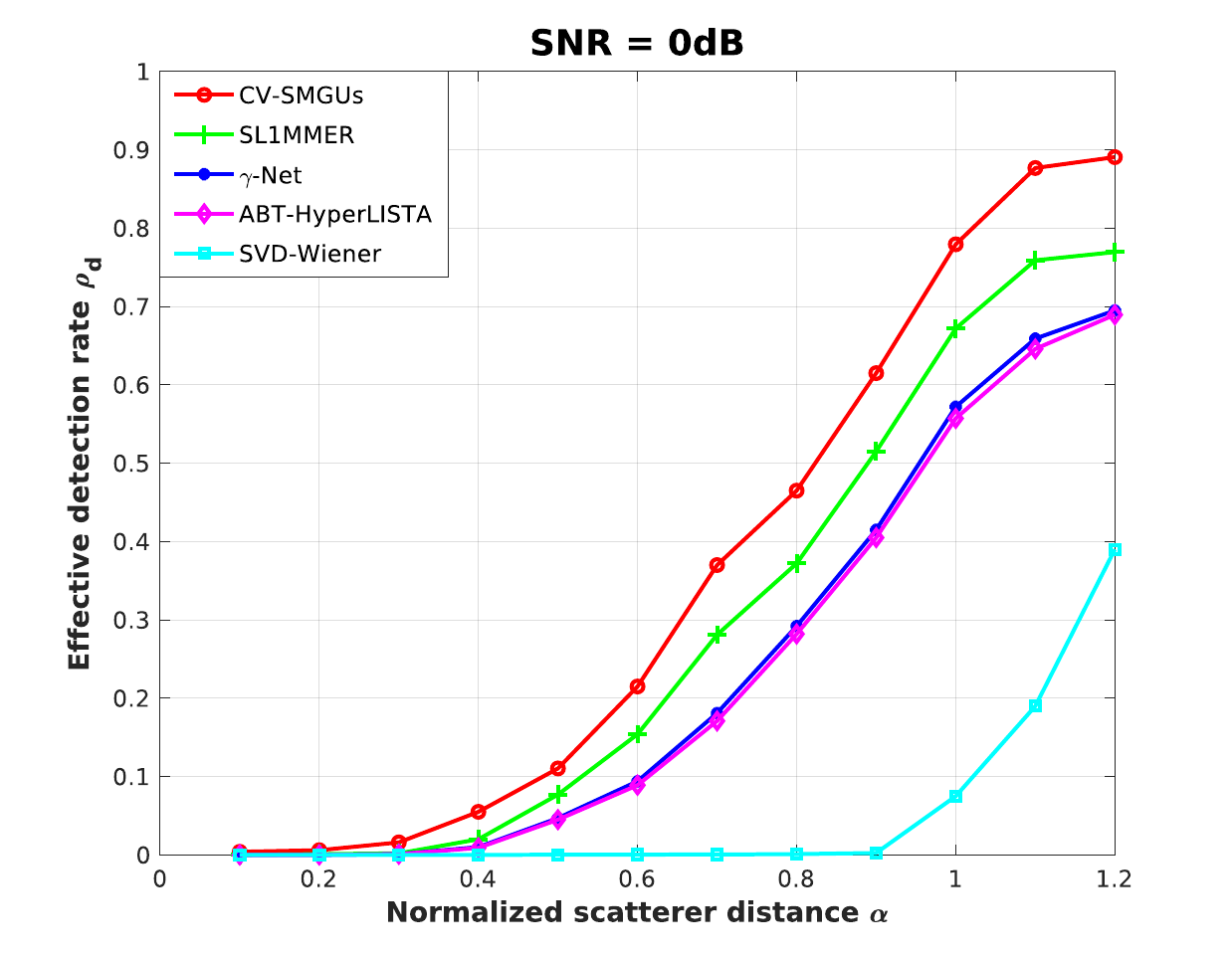}
    \caption*{(a)}
    \end{minipage}
    \begin{minipage}[t]{0.49\linewidth}
    \includegraphics[width=0.98\textwidth]{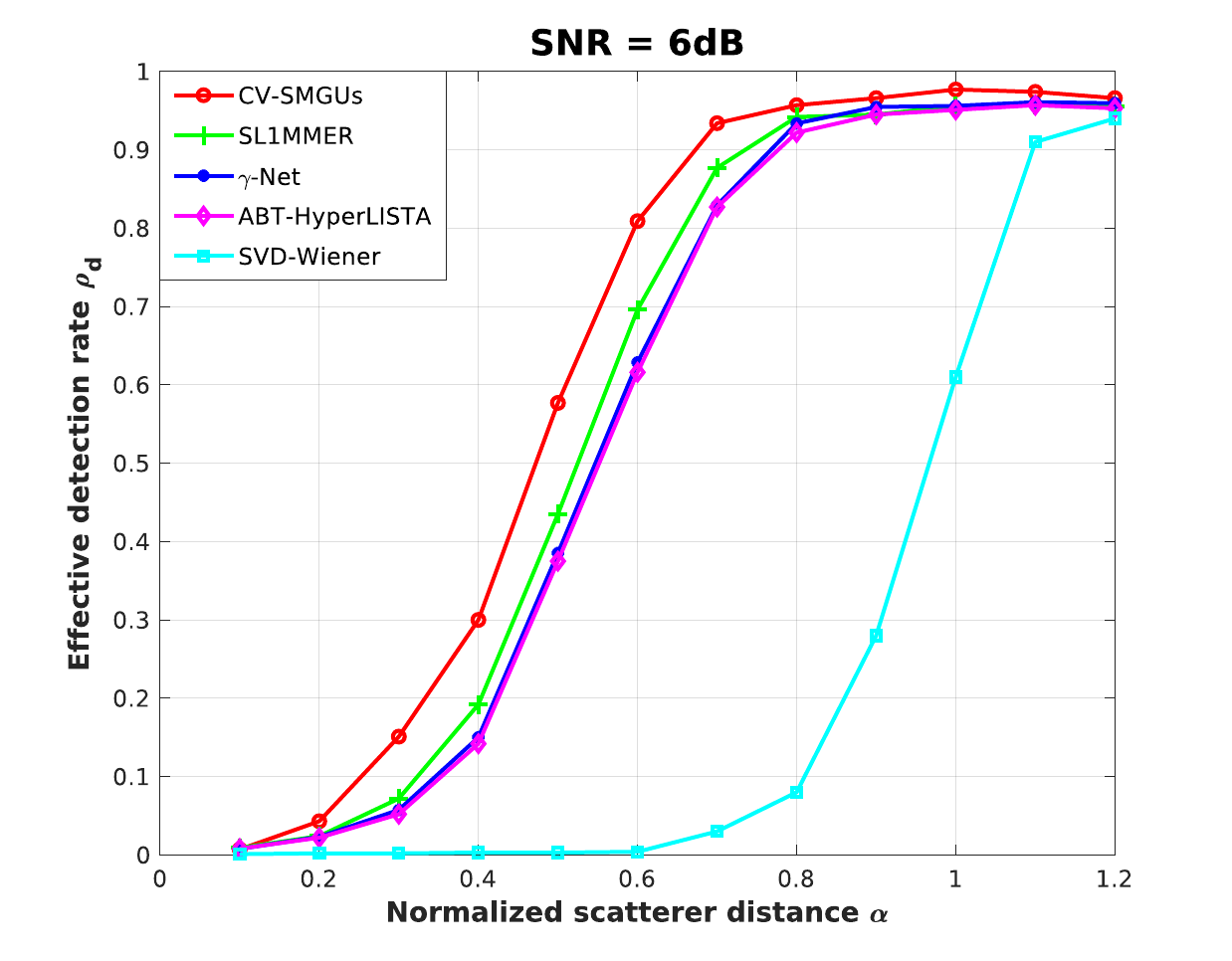}
    \caption*{(b)}
    \end{minipage}
    \caption{Detection rate $P_d$ as a function of the normalized elevation distance between the simulated facade and ground with SNR = 0 dB and 6 dB, N = 25, and phase difference $\triangle \phi = 0$ (worst case) under 0.2 million Monte Carlo trials.}
    \label{fig:detec_rate}
\end{figure*}

The comparison was first based on the effective detection rate. Two different scenarios were taken into consideration: SNR $\in \{0, 6 \}$dB, which represents a noisy case, and a regular case with a typical SNR level in a high-resolution spaceborne SAR image. The comparison results are demonstrated in Fig. \ref{fig:detec_rate}. At each discrete normalized elevation distance, 0.2 million Monte Carlo trials with an identical phase and amplitude, which represents the worst case \cite{Zhu2012Super-Resolution} in TomoSAR inversion, were simulated. The deep learning-based algorithms $\boldsymbol{\gamma}$-Net and CV-SMGUs were built with 12 and 6 hidden layers, respectively. The training followed the same training strategy introduced in \cite{gamma-net_QK} \cite{SMGU_QK} and was carried out using a single NVIDIA RTX2080 GPU. For HyperLISTA-ABT, the training involved analytical weight optimization and determining the hyperparameters via the grid search method. The number of iterations in HyperLISTA-ABT was set as 15. 

From the comparison results, we can see that all the methods except the conventional spectral estimator SVD-Wiener showed a great super-resolution power. The proposed HyperLISTA-ABT delivered almost the same super-resolution ability as $\boldsymbol{\gamma}$-Net and approached the performance of SL1MMER in both scenarios. When focusing solely on the effective detection rate, it was challenging to proclaim a clear advantage of the proposed HyperLISTA-ABT method over the existing state-of-the-art approaches. In fact, when comparing it to CV-SMGUs, we could observe a slight underperformance. However, all the state-of-the-art methods come with a relatively high computational cost. Both $\boldsymbol{\gamma}$-Net and CV-SMGU require tailored training according to the baseline distribution of the stack. SL1MMER is a model-based algorithm, thus needs no training, yet requires significantly computational time for solving the L1-norm minimization. 

We tested and recorded the time consumption of different algorithms for processing the 0.2 million Monte Carlo trials as well as the requirements for training data. The results are summarized in Table \ref{tab:comp_RNNs}. To clarify, all inference was conducted using a local CPU for a fair comparison. As can be seen in Table \ref{tab:comp_RNNs}, it took about ten hours for the deep learning-based algorithms to process 0.2 million Monte Carlo trials, which was predominantly the training time. In addition, a large amount of training samples was essential as well. For SL1MMER, it took about 20 hours for the processing since the iterative second-order optimization is computationally expensive. Further inspecting the table, we can see that HyperLISTA-ABT showed similar efficiency in the inference as the other deep learning-based algorithms. However, HyperLISTA-ABT required no training data and it took much less time for the training. In total, HyperLISTA-ABT speeded up the processing by about one order of magnitude compared to the other algorithms tested in the experiment. 

\begin{table*}[h!]
\centering
\caption{Comparison of the number of required training samples and time consumption for processing 0.2 million Monte Carlo trials with each algorithm. The training time of HyperLISTA-ABT indicates the combined duration of both the analytic weight optimization process and the tuning of hyperparameters. It provides a measure of the overall time required for these essential steps.}
\resizebox{\linewidth}{!}{
\begin{tabular}{lllllll}
   \toprule
\textbf{Algorithm}        & number of training samples  & training time & inference time & total time consumption & transferability \\
      \midrule % <-- Midrule here
CV-SMGUs & 4 million & $\approx 10$ hours & $\approx 0.25$ h & $\approx 10$ h & low \\
$\boldsymbol{\gamma}$-Net & 3 million & $\approx 8$ hours & $\approx 0.2$ h & $\approx 8$ h & low\\
SL1MMER & - & - & $\approx 20$ h & $\approx 20$ h & high \\
HyperLISTA-ABT & - & $\approx 0.5$ hour & $\approx 0.25$ h & $\approx 1$ h & medium\\
\bottomrule
\end{tabular}}
\label{tab:comp_RNNs}
\end{table*}

Upon evaluating the performance and efficiency, it was observed that HyperLISTA-ABT achieved comparable performance to existing state-of-the-art methods while significantly improving the computational efficiency by approximately one order of magnitude. This is especially advantageous in the multi-component D-TomoSAR case. The application of the aforementioned deep learning-based algorithms and SL1MMER are very limited in the D-TomoSAR case due to the need of time-consuming model training and the heavy computational expense. On the contrary, the application of HyperLISTA-ABT can be easily extended to computationally efficient D-TomoSAR processing.  Therefore, HyperLISTA-ABT is a more applicable approach for the large-scale processing of real data.

Furthermore, HyperLISTA-ABT demonstrates superior transferability compared to deep learning-based algorithms. Deep learning models are typically trained to fit specific baseline configurations, such as a fixed number of SAR acquisitions and a specific baseline distribution. While they may exhibit satisfactory generalizability to small baseline discrepancies \cite{gamma-net_QK} \cite{SMGU_QK}, directly applying a trained deep learning model to a new data stack with a different number of acquisitions or a completely different baseline distribution is not feasible. In such cases, time-consuming retraining of the model becomes necessary, resulting in low transferability.

In contrast, HyperLISTA-ABT offers better transferability. Although it requires analytical optimization of the weight matrix for each new data stack, the efficiency of the analytical optimization process allows for scalability and improved transferability. This finding highlights the potential of HyperLISTA-ABT in enabling global urban mapping using TomoSAR, as it can be effectively applied to diverse data stacks with varying acquisition configurations and baseline distributions.

\section{Real data experiment}
\subsection{Bellagio hotel}
\begin{figure}[h]
    \centering
    \includegraphics[width=0.49\textwidth]{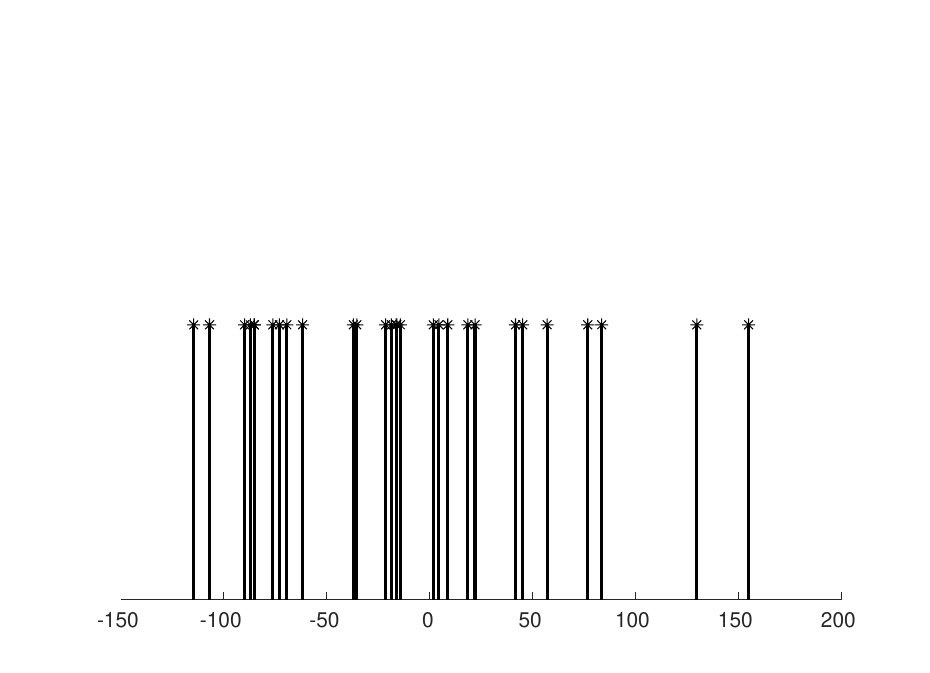}
    \caption{Effective baselines of the 29 TerraSAR-X high-resolution spotlight images.}
    \label{fig:base_dis}
\end{figure}

\begin{figure*}[h]
    \begin{minipage}[t]{0.49\linewidth}
    \centering
    \includegraphics[height=7.5cm]{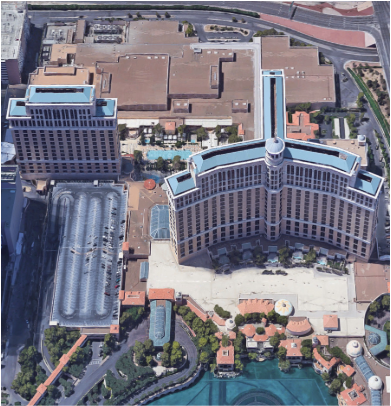}
    \caption*{(a)}
    \end{minipage}
    \begin{minipage}[t]{0.49\linewidth}
    \centering
    \includegraphics[height=7.5cm]{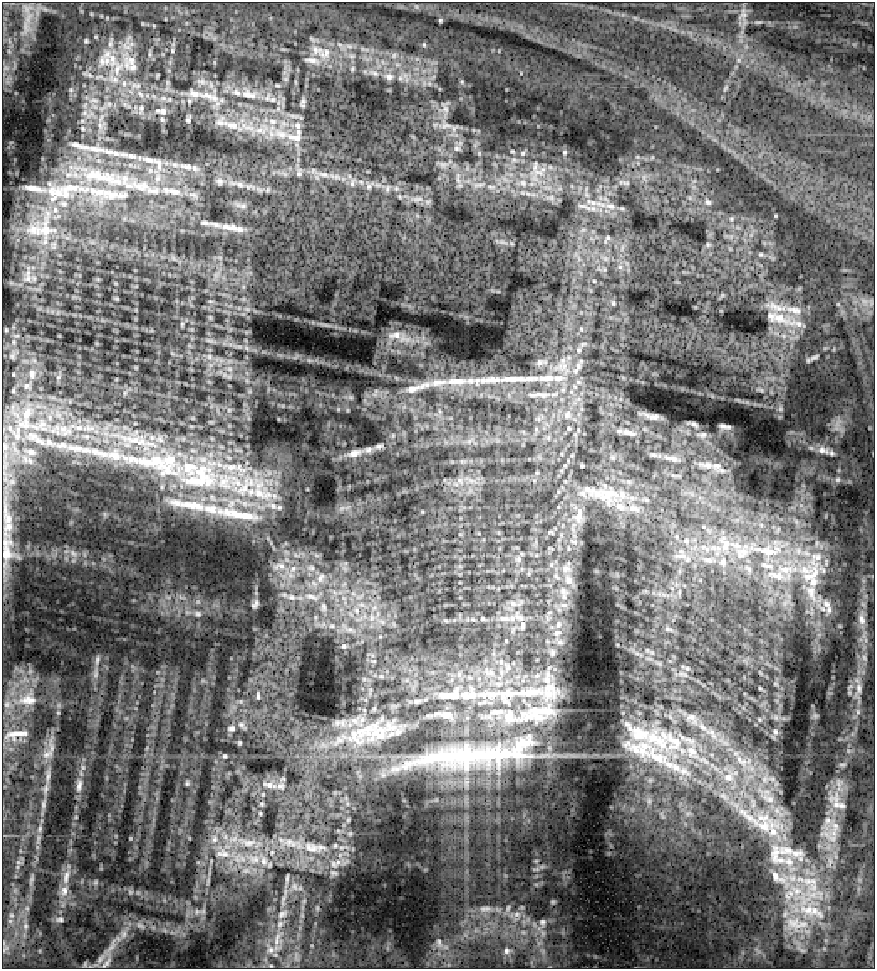}
    \caption*{(b)}
    \end{minipage}
    \caption{Test site. (a): optical image from Google Earth, (b): SAR mean intensity image}
    \label{fig:test_area}
\end{figure*}

In this real data experiment, due to the fact that there was no available ground truth, we purposely used the same data as in \cite{zhu2012demo} so that we can compare our results to the results obtained with SL1MMER. The datastack was composed of 29 TerraSAR-X high-resolution spotlight images covering the Bellagio Hotel in Las Vegas, whose baseline distribution is illustrated in Fig. \ref{fig:base_dis}. The slant-range resolution was 0.6m and the azimuth resolution was 1.1m. The elevation aperture size of about 270m resulted in the inherent elevation resolution $\rho_s$ to be about 40m, i.e. approximately 20m resolution in height since the incidence angle here was $31.8^{\circ}$. An optical image and the SAR mean intensity image of the test site are shown in Fig. \ref{fig:test_area} 

As for the D-TomoSAR system model, a time wrap operation assuming only sinusoidal seasonal motion was adopted as in \cite{zhu_4D_tomosar} because no long-term linear motion was observed during the acquisition period of the test area. 

\begin{figure*}[h]
    \includegraphics[width = 0.98\textwidth]{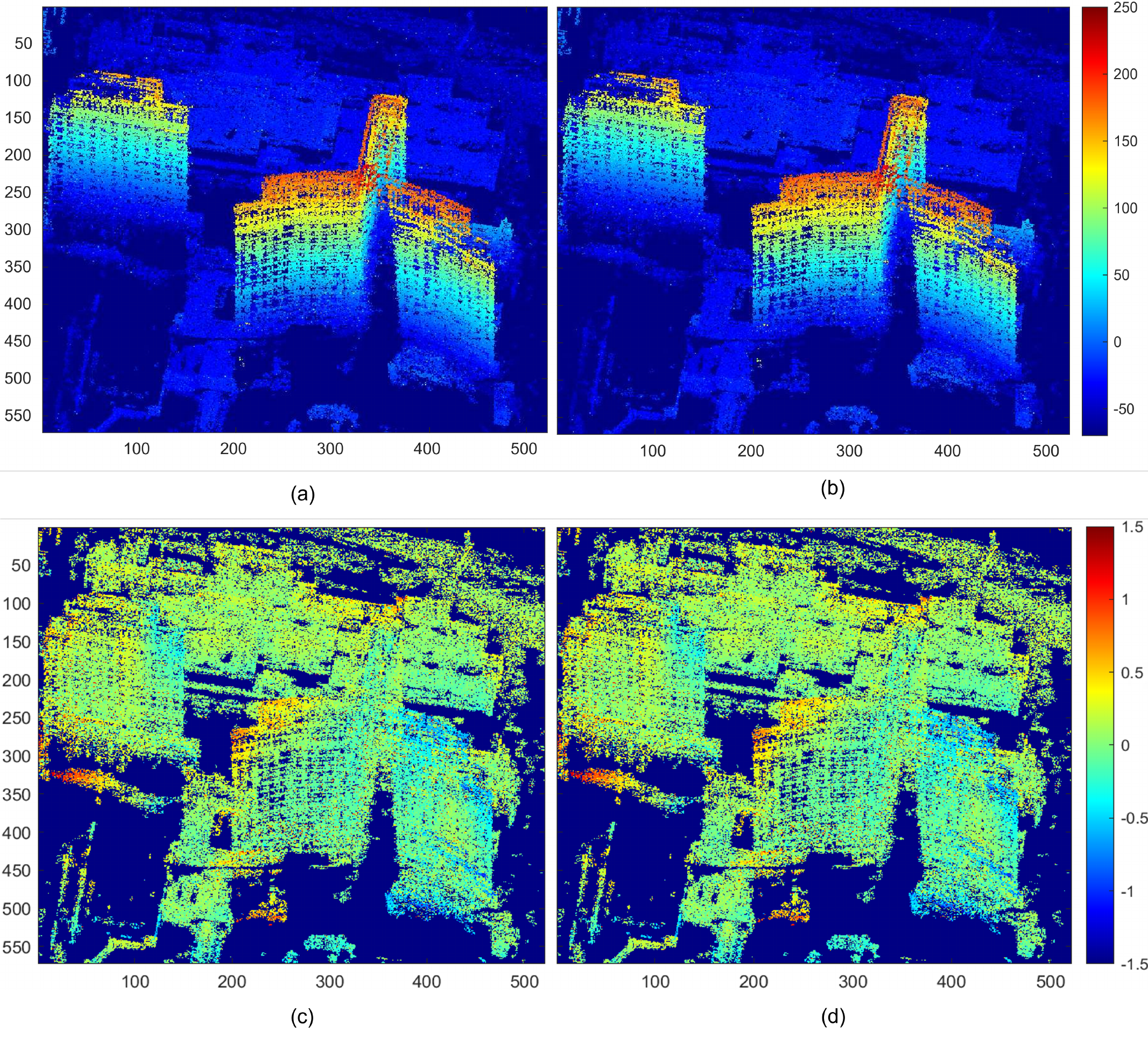}
    \caption{Color-coded reconstruction results of the test site. (a) Elevation estimates using HyperLISTA-ABT in meters, (b) elevation estimates using SL1MMER in meters, (c) estimated amplitude of seasonal motion using HyperLISTA-ABT in centimeters, (d) estimated amplitude of seasonal motion using SL1MMER in centimeters.}
    \label{fig:recon_bellagio}
\end{figure*}

\begin{figure*}[h]
    \includegraphics[width = 0.98\textwidth]{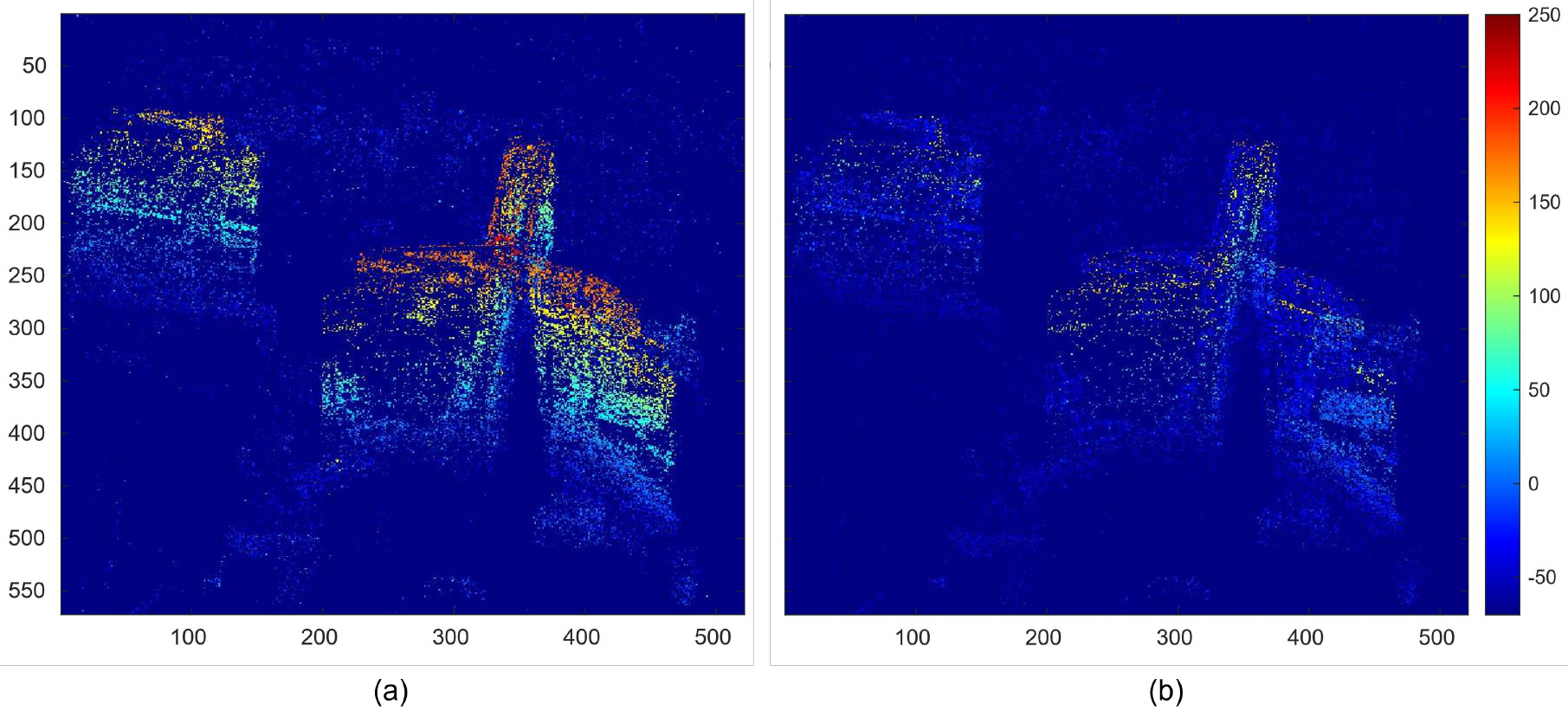}
    \caption{Color-coded elevation estimates of the top and bottom layers of detected double scatterers using HyperLISTA-ABT. (a) Top layer, mostly caused by reflections from the building roof and facade, (b) bottom layer, mostly caused by reflections from low infrastructures and the ground.}
    \label{fig:sep_layers}
\end{figure*}

In Fig. \ref{fig:recon_bellagio}, we compare the estimated elevation and amplitude of the seasonal motion of the detected single scatterers and the top layer of the detected double scatterers. From Fig. \ref{fig:recon_bellagio}(a), we can see a smooth gradation of the elevation estimates from the building bottom top, which suggests a reasonable elevation estimation by HyperLISTA-ABT. Moreover, we can see that there is no significant difference between the results of HyperLISTA-ABT and SL1MMER, implying that HyperLISTA-ABT had similar performance to SL1MMER. In addition, Fig. \ref{fig:sep_layers} shows the layover separation ability of HyperLISTA-ABT. As can be seen, the two layers of double scatterers were detected and separated by HyperLISTA-ABT. The top layer was mainly caused by signals from the roof and facade of the high rise building while the bottom layer was caused by signals from the ground structures. 

We also conducted some numerical comparisons of both algorithms. First, we compared the percentage of pixels detected as zero, one, and two scatterers by both algorithms. Compared to SL1MMER, we found that HyperLISTA-ABT detected more pixels as coherent sactterers. This does not necessarily mean that HyperLISTA-ABT had a better detection ability since there was no ground truth. We believe HyperLISTA-ABT detected more scatterers because HyperLISTA-ABT tends to maintain weak signals, which could be reflections of dark scatterers but also outliers caused by noise interference. The false detection of noise as coherent scatterers causes a speckle-like noise in the reconstruction result. Model order selection and post-processing  techniques like spatial filtering can further mitigate such outliers. 

\begin{figure}[h]
    \centering
    \includegraphics[width=0.49\textwidth]{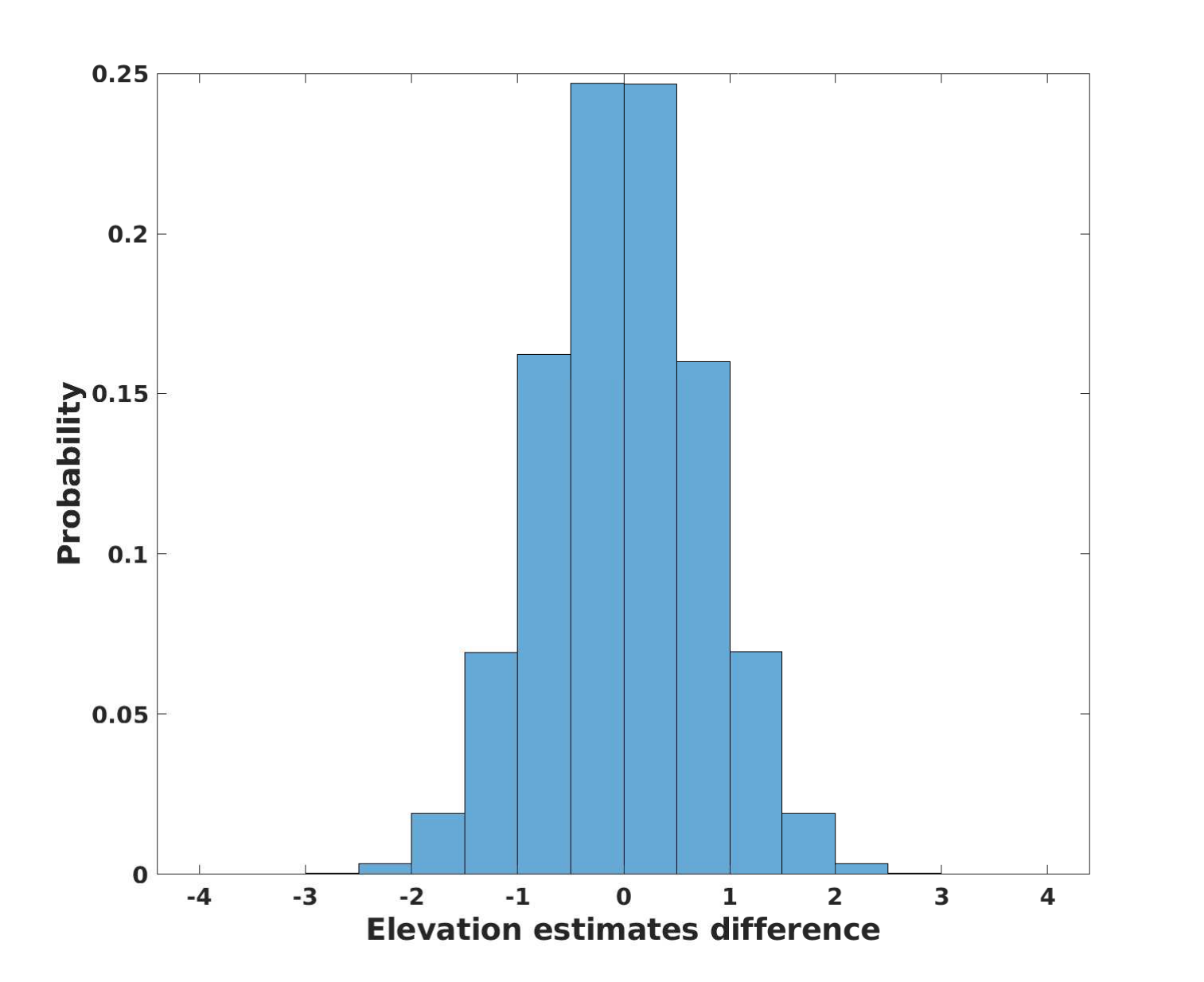}
    \caption{Histogram of elevation estimates differences between HyperLISTA-ABT and SL1MMER.}
    \label{fig:ele_dif}
\end{figure}

For further evaluation, we compared the elevation estimates differences of scatterers detected by both algorithms. A histogram of the elevation estimates differences is shown in Fig. \ref{fig:ele_dif}. It can be observed that most of the elevation estimates differences were within 1 meter. This observation indicates that both algorithms yielded comparable results in terms of elevation estimation, instilling confidence in their reliability and reasonableness. Furthermore, this similarity in estimation accuracy suggests that HyperLISTA-ABT performed on par with SL1MMER. Moreover, it is worth mentioning that it took more than three weeks for SL1MMER to finish the D-TomoSAR processing over the test site, whereas it only took HyperLISTA-ABT several hours to complete the processing.

\begin{table}[h]
    \centering
    \caption{Percentage of scatterers detection for the two algorithms.}
    \begin{tabular}{p{0.15\textwidth}| p{0.08\textwidth} p{0.08\textwidth} p{0.08\textwidth}}
    \toprule
    \multirow{2}{*}{Algorithm}  & \multicolumn{3}{c}{Percentage of detection as} \\
     & 0 scatterer & 1 scatterer & 2 scatterers \\
     \midrule
     HyperLISTA-ABT & 48.48 $\%$ & 44.09 $\%$ & 7.43 $\%$   \\
     SL1MMER & 49.41 $\%$ & 43.63 $\%$ & 6.96 $\%$  \\
     \bottomrule
    \end{tabular}
    \label{tab:n_scatterer}
\end{table}

\subsection{Large demonstration}
\begin{figure*}[h]
    \centering
    \includegraphics[height=8cm]{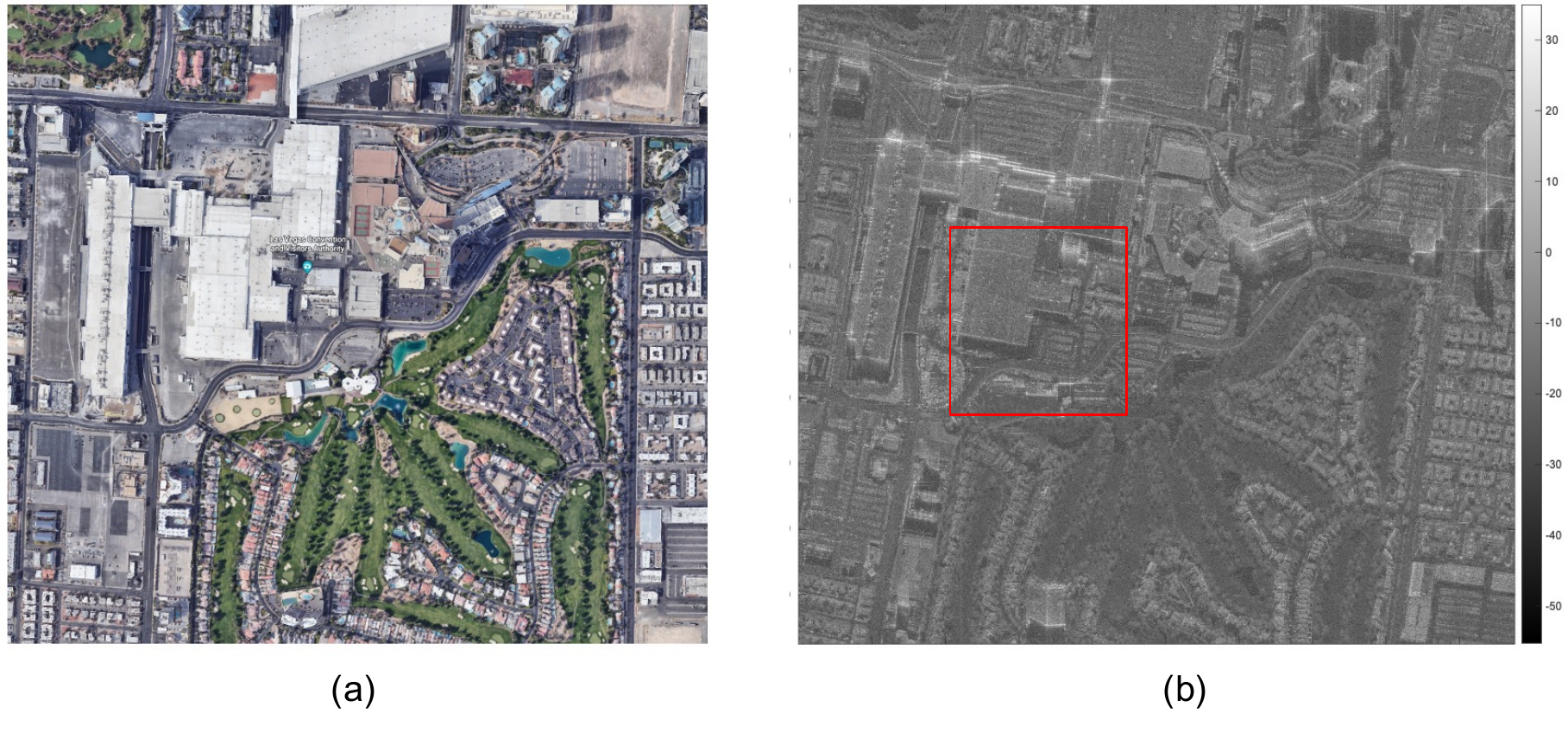}
    \caption{Demonstration of the large test area. (a) Optical image from Google Earth, (b) SAR mean intensity map in dB. The red box in (b) indicates the area undergoing subsidence.}
    \label{fig:large_area}
\end{figure*}

In this section, we applied HyperLISTA-ABT to TerraSAR-X high-resolution spotlight data over a large area surrounding the convention center in Las Vegas. The stack was composed of 29 acquisitions covering a time period from July 2009 to June 2010, during which the test area was undergoing a pronounced subsidence centered at the convention center. Therefore, the acquisitions were characterized by a multi-component nonlinear motion combining linear and thermal-dilation-induced seasonal motion. Fig. \ref{fig:large_area} shows us an optical image and the SAR mean intensity map. The red box indicates the "epiccenter" undergoing subsidence around the convention center. 

\begin{figure*}[h]
    \centering
    \includegraphics[width=0.98\textwidth]{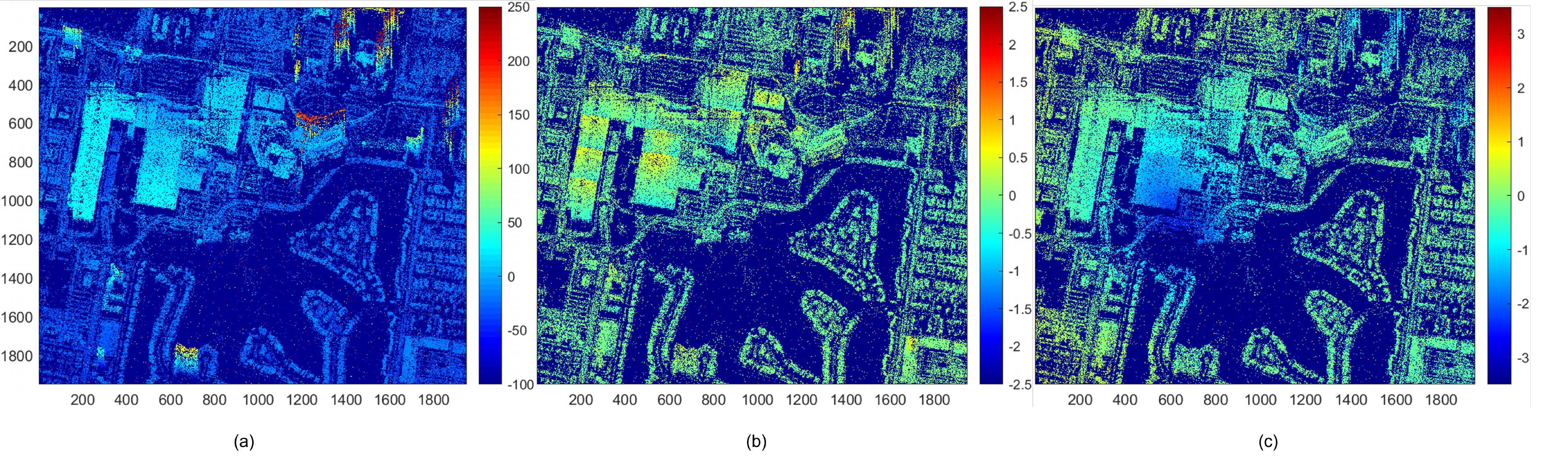}
    \caption{Demonstration of color-coded elevation estimates and estimated amplitude of multi-component motion. (a) Elevation estimates in meters, (b) estimated amplitude of seasonal motion in centimeters, (c) estimated amplitude of linear motion in centimeters/year.}
    \label{fig:large_area_process}
\end{figure*}

Fig. \ref{fig:large_area_process} illustrates us the reconstructed elevation estimates as well as the estimated amplitude maps of the two different motions. As we can see from the surface model generated from the elevation estimates in Fig. \ref{fig:large_area_process}(a), we can capture the shapes of individual buildings and the surrounding infrastructures, like roads, at a detailed level. In addition, Fig. \ref{fig:large_area_process}(b) shows that clear deformation caused by seasonal motion can be observed in the metallic building structures since they were affected by thermal dilation more seriously compared to surrounding infrastructures. Furthermore, as illustrated in Fig. \ref{fig:large_area_process}(c), it could be observed that the magnitude of the linear subsidence increased as the scatterer getting closer to the "epicenter". These results are consistent with the fact, thus validating the effectiveness of HyperLISTA-ABT for multi-component nonlinear motion estimation and giving confidence that HyperLISTA-ABT can be applied in large-scale D-TomoSAR processing.

\section{Conclusion}
This paper proposes HyperLISTA-ABT to address the gap in applying deep neural networks for solving D-TomoSAR inversion. Unlike traditional methods that learn weights directly from data, HyperLISTA-ABT computes the weights with an analytical optimization technique by minimizing generalized mutual coherence. Additionally, HyperLISTA-ABT introduces an adaptive blockwise thresholding scheme that applies block coordinate techniques to accelerate the algorithm. Moreover, it conducts thresholding in local blocks to retain weak expressions of reflection from dark scatterers and consider more local features. Laboratory experiments for 3D reconstruction confirmed the efficiency of HyperLISTA-ABT in estimation. Moreover, tests on real data over a large area demonstrated that HyperLISTA-ABT can reconstruct high-quality 4D point clouds, making it an efficient and accurate algorithm for future large-scale or even global D-TomoSAR processing.

% or


\appendix
\section{Appendixes}
\subsection{Efficient algorithm to calculate weight analytically}
As discussed in \cite{liu2018alista}, it is difficult to solve Eq. (\ref{eq:min_coh}) directly and Eq. (\ref{eq:min_coh}) can be reformulated as minimizing the Frobenius norm of $\mathbf{W}^H \mathbf{R}$ over a linear constraint. Defining $\mathbf{W} = \mathbf{G}^H \mathbf{GR}$ ($\mathbf{G} \in \mathbb{C} ^{N \times N}$ is named as the Gram matrix), the minimization of the Frobenius norm reads:
\begin{equation}
    \min _G\left\|R^H \boldsymbol{G}^H \boldsymbol{G R}-\boldsymbol{I}\right\|_F^2, \quad \text { s.t. } \operatorname{diag}\left(\boldsymbol{R}^H \boldsymbol{G}^H \boldsymbol{G} \boldsymbol{R}\right)=\mathbf{1} .
    \label{eq:min_G}
\end{equation}
However, it is hard to handle the constraint in the above problem (\ref{eq:min_G}). As a solution, a matrix $\mathbf{D = G R} \in \mathbb{C}^{N \times L}$ is introduced and we use the following method as an alternative:
\begin{equation}
    \min _{\boldsymbol{G}, \boldsymbol{D}}\left\|\boldsymbol{D}^T \boldsymbol{D}-\boldsymbol{I}\right\|_F^2+\frac{1}{\alpha}\|\boldsymbol{D}-\boldsymbol{G} \boldsymbol{R}\|_F^2 \text {, s.t. } \operatorname{diag}\left(\boldsymbol{D}^T \boldsymbol{D}\right)=\mathbf{1} \text {. }
    \label{eq:min_W}
\end{equation}
With a proper $\alpha>0$, the solution to Eq. (\ref{eq:min_W}) approximates Eq. (\ref{eq:min_G}) and we obtain the optimized weights accordingly. The steps for solving the optimization problem (\ref{eq:min_W}) are described as follows.

First, $\mathbf{G}$ is fixed and we update $\mathbf{D}$ with the projected gradient descent (PGD):
\begin{align}
    \boldsymbol{D} \leftarrow \mathcal{P}\left(\boldsymbol{D}-\zeta \boldsymbol{D}\left(\boldsymbol{D}^H \boldsymbol{D}-\boldsymbol{I}\right)-\frac{\zeta}{\alpha}(\boldsymbol{D}-\boldsymbol{G} \boldsymbol{R})\right)
    \label{eq:opt_D}
\end{align}
where $\mathcal{P}$ denotes the projection operator on the constraint $\mathrm{diag}\left(\boldsymbol{D}^T \boldsymbol{D}\right)=\mathbf{1}$, so that each column of $\mathbf{D}$ will be normalized; and $\zeta$ is the stepsize. Hereafter, we fix $D$ and update the minimizer of $G$ with
\begin{equation}
    \boldsymbol{G} \leftarrow \mathbf{D R}^{+}
    \label{eq:opt_G}
\end{equation}
where $\mathbf{R}^{+}$ represents the Moore-Penrose pseudoinverse of the steering matrix $\mathbf{R}$. Then, we repeat the procedure until $\mathbf{D} \approx \mathbf{G R}$. The whole algorithm is summarized in \textbf{Algorithm 2}.
\begin{algorithm}[h]
\caption{Efficient algorithm for analytical weight optimization}
\label{table:algorithm}
\begin{algorithmic}
    \STATE \textbf{Input:} the steering matrix $\mathbf{R}$\\
    \STATE \textbf{Init:} $\mathbf{D=R}$, $\mathbf{G=I}$, $\zeta=\alpha=0.1$\\
    \STATE \textbf{for} iter$=1,2,\cdots$ until \textbf{convergence do}  \\
    \STATE \qquad update $\mathbf{D}$ with (\ref{eq:opt_D}) \\
    \STATE \qquad update $\mathbf{G}$ with (\ref{eq:opt_G})\\
    \STATE \qquad Compute $f_1=\left\|\boldsymbol{D}^H \boldsymbol{D}-\boldsymbol{I}\right\|_F^2$\\
    \STATE \qquad Compute $f_2=\left\|(\boldsymbol{G} \boldsymbol{R})^H \boldsymbol{G} \boldsymbol{A}-\boldsymbol{I}\right\|_F^2$\\
    \STATE \qquad \textbf{if} two consecutive $f_1$s are close enough \textbf{then} \\
    \STATE \qquad \qquad $\zeta=0.1\zeta$ \\ 
    \STATE \qquad \qquad $\alpha=0.1\alpha$ \\ 
    \STATE \qquad \qquad \textbf{if} $f_1$ and $f_2$ are close enough \textbf{then} \\
    \STATE \qquad \qquad \qquad \textbf{break}
    \STATE \qquad \qquad \textbf{end}
    \STATE \qquad \textbf{end}
    \STATE \textbf{end}
    \STATE \textbf{Output:} $\mathbf{W} = \mathbf{G}^H \mathbf{GR}$
 \end{algorithmic}
\end{algorithm}

\ifCLASSOPTIONcaptionsoff
  \newpage
\fi



%\bibliographystyle{IEEEtran}
%\clearpage
\bibliographystyle{IEEEtran}
\bibliography{ref}
%\begin{thebibliography}{1}


%\end{thebibliography}

\begin{IEEEbiography}[{\includegraphics[width=1in,height=1.25in,clip,keepaspectratio]{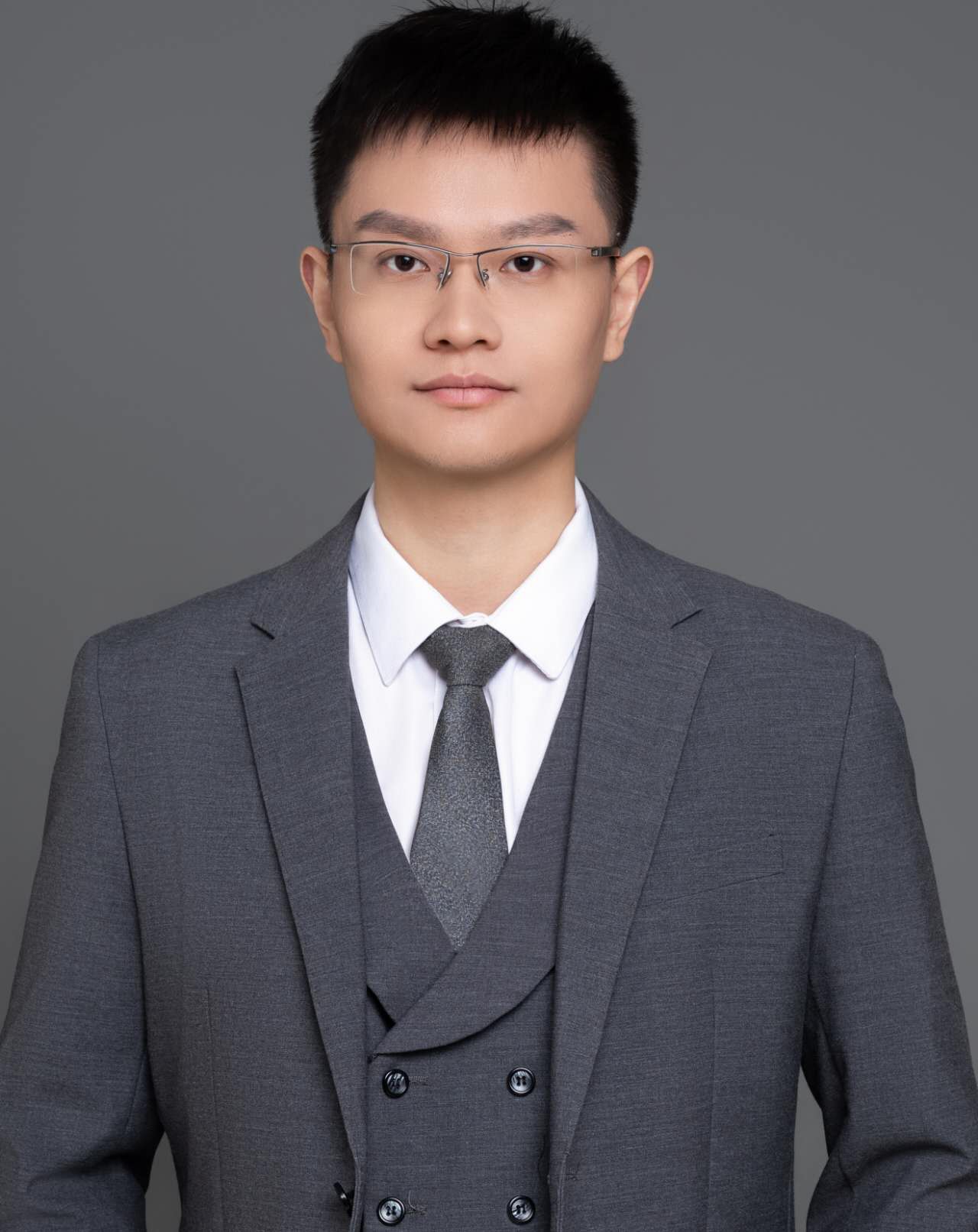}}]{Kun Qian}
received double B.Sc. degree in Remote Sensing and Information Engineering from Wuhan University, Wuhan, China and Aerospace Engineering and Geodesy from University of Stuttgart, Stuttgart, Germany, respectively, in 2016, and M.Sc. degree in Aerospace Engineering and Geodesy from University of Stuttgart, Stuttgart, Germany in 2018. He is pursuing the Ph.D. degree with Data Science in Earth Observation, Technical Unversity of Munich, Munich, Germany. His research focus includes data-driven methods, deep unfolding algorithms and their application in multi-baseline SAR tomography.
\end{IEEEbiography}

\begin{IEEEbiography}[{\includegraphics[width=1in,height=1.25in,clip,keepaspectratio]{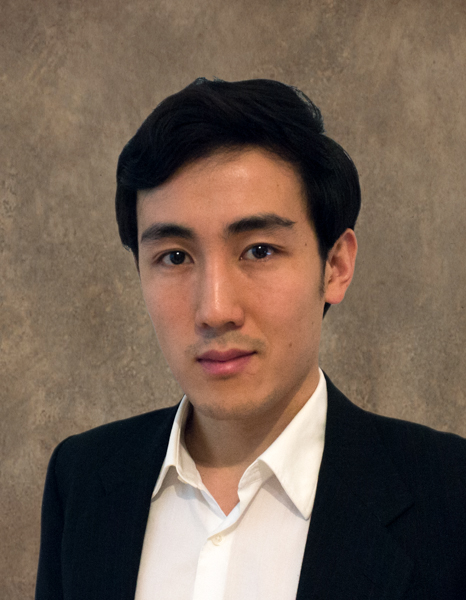}}]{Yuanyuan Wang}(S'08--M'15)
received the B.Eng. degree (Hons.) in Electrical Engineering from The Hong Kong Polytechnic University, Hong Kong, China in 2008, and the M.Sc. and Dr.-Ing. degree from the Technical University of Munich, Munich, Germany, in 2010 and 2015, respectively. In June and July 2014, he was a Guest Scientist with the Institute of Visual Computing, ETH Zürich, Zürich, Switzerland. He is currently a Guest Professor at the German Interantional AI Future Lab: AI4EO, at the Technical University of Munich. He is also with the Department of EO Data Science, in the Remote Sensing Technology Institute of the German Aerospace Center, where he leads the working group Big SAR Data. His research interests include optimal and robust parameters estimation in multibaseline InSAR techniques, multisensor fusion algorithms of SAR and optical data, nonlinear optimization with complex numbers, machine learning in SAR, uncertainty quantification and mitigation in machine learning, and high-performance computing for big data. Dr. Wang was one of the best reviewers of the IEEE Transactions ON Geoscience and Remote Sensing in 2016. He is a Member of the IEEE.
\end{IEEEbiography}

\begin{IEEEbiography}[{\includegraphics[width=1in,height=1.25in,clip,keepaspectratio]{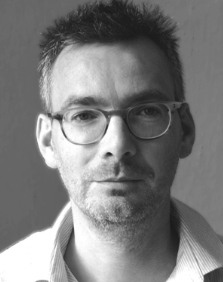}}]{Peter Jung} (Member IEEE) received the Dipl.-Phys. degree in high energy physics from Humboldt University, Berlin, Germany, in 2000, in cooperation with DESY Hamburg, and the Dr.-rer.nat (Ph.D.) degree in Weyl–Heisenberg representations in communication theory with the Technical University of Berlin (TUB), Germany, in 2007. Since 2001, he has been with the Department of Broadband Mobile Communication Networks, Fraunhofer Institute for
Telecommunications, Heinrich-Hertz-Institut (HHI), and since 2004 with the
Fraunhofer German-Sino Laboratory for Mobile Communications. He is
currently working under DFG grants at TUB in the field of signal processing,
information and communication theory, and data science. He is also a
Visiting Professor with TU Munich and associated with the Munich AI
Future Laboratory (AI4EO). His current research interests include the area
compressed sensing, machine learning, time–frequency analysis, dimension
reduction, and randomized algorithms. He is giving lectures in compressed
sensing, estimation theory and inverse problems. He is also a member of
VDE/ITG.
\end{IEEEbiography}

\begin{IEEEbiography}[{\includegraphics[width=1in,height=1.25in,clip,keepaspectratio]{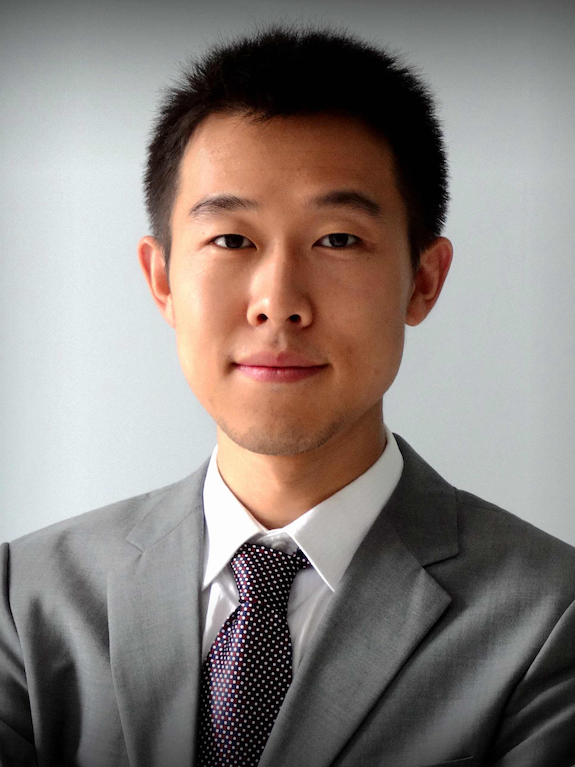}}]{Yilei Shi} (Member, IEEE) received the Dipl.-Ing. degree in mechanical engineering and the Dr.-Ing. degree in signal processing from Technische Universität München (TUM), Munich, Germany, in 2010 and 2019, respectively. He is a Senior Scientist with the Chair of Remote Sensing Technology, TUM. His research interests include fast solver and parallel computing for large scale problems, high-performance computing and computational intelligence, advanced methods on SAR and InSAR processing, machine learning and deep learning for variety of data sources, such as SAR, optical images, and medical images, and PDE-related numerical modeling and computing.
\end{IEEEbiography}

\begin{IEEEbiography}[{\includegraphics[width=1in,height=1.25in,clip,keepaspectratio]{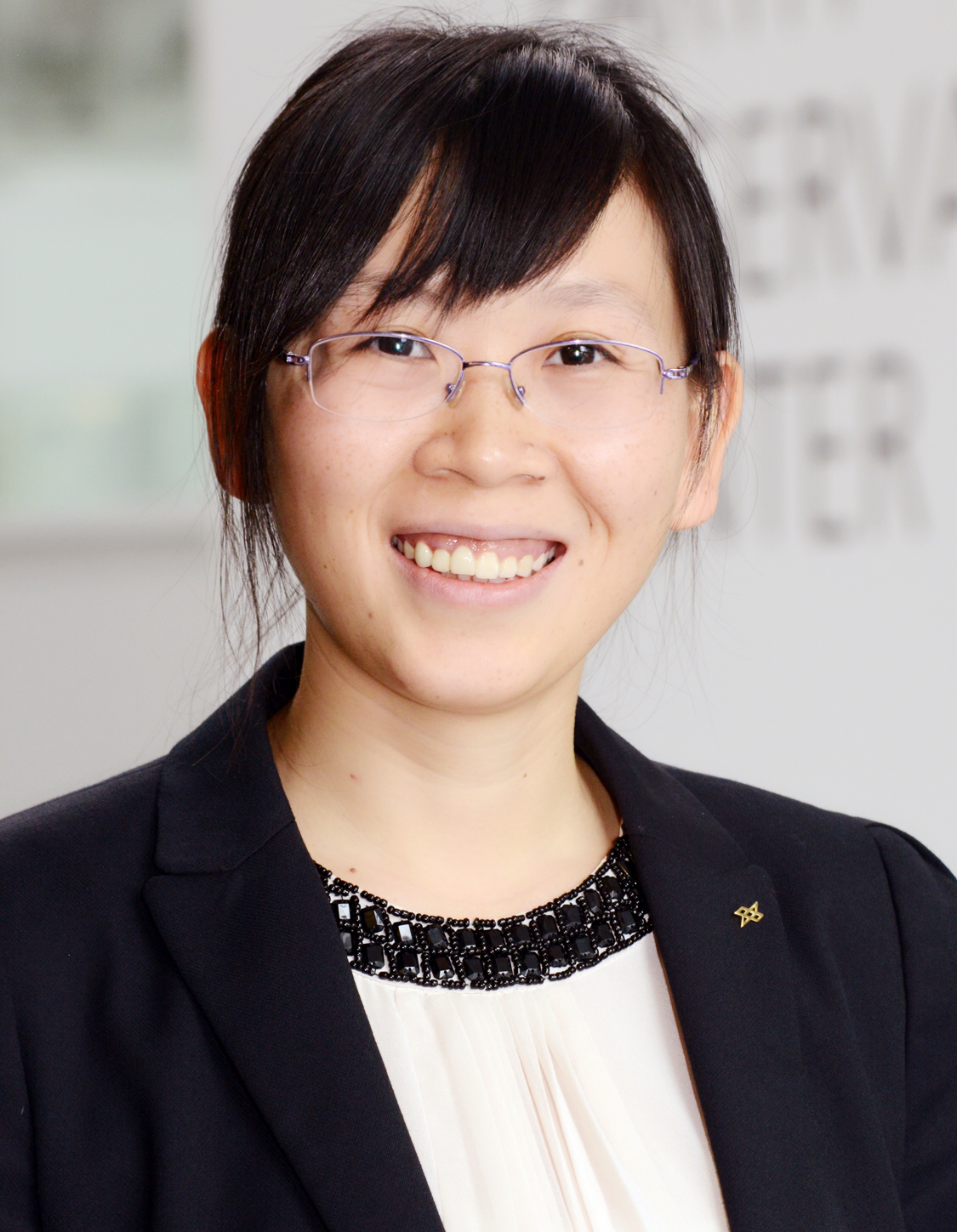}}]{Xiao Xiang Zhu}(S'10--M'12--SM'14--F'21) received the Master (M.Sc.) degree, her doctor of engineering (Dr.-Ing.) degree and her “Habilitation” in the field of signal processing from Technical University of Munich (TUM), Munich, Germany, in 2008, 2011 and 2013, respectively.
\par
She is currently the Chair Professor for Data Science in Earth Observation (former: Signal Processing in Earth Observation) at Technical University of Munich (TUM) and was the Founding Head of the Department ``EO Data Science'' at the Remote Sensing Technology Institute, German Aerospace Center (DLR). Since 2019, Zhu is a co-coordinator of the Munich Data Science Research School (www.mu-ds.de). Since 2019 She also heads the Helmholtz Artificial Intelligence -- Research Field ``Aeronautics, Space and Transport". Since May 2020, she is the director of the international future AI lab "AI4EO -- Artificial Intelligence for Earth Observation: Reasoning, Uncertainties, Ethics and Beyond", Munich, Germany. Since October 2020, she also serves as a co-director of the Munich Data Science Institute (MDSI), TUM. Prof. Zhu was a guest scientist or visiting professor at the Italian National Research Council (CNR-IREA), Naples, Italy, Fudan University, Shanghai, China, the University  of Tokyo, Tokyo, Japan and University of California, Los Angeles, United States in 2009, 2014, 2015 and 2016, respectively. She is currently a visiting AI professor at ESA's Phi-lab. Her main research interests are remote sensing and Earth observation, signal processing, machine learning and data science, with a special application focus on global urban mapping.

Dr. Zhu is a member of young academy (Junge Akademie/Junges Kolleg) at the Berlin-Brandenburg Academy of Sciences and Humanities and the German National  Academy of Sciences Leopoldina and the Bavarian Academy of Sciences and Humanities. She serves in the scientific advisory board in several research organizations, among others the German Research Center for Geosciences (GFZ) and Potsdam Institute for Climate Impact Research (PIK). She is an associate Editor of IEEE Transactions on Geoscience and Remote Sensing and serves as the area editor responsible for special issues of IEEE Signal Processing Magazine. She is a Fellow of IEEE.
\end{IEEEbiography}


%\vfill


\end{document}